\RequirePackage{fix-cm}
\documentclass[smallcondensed]{svjour3}     
\usepackage{graphicx}
\usepackage{subcaption}
\usepackage{amssymb}
\usepackage{mathtools}
\usepackage[breaklinks=true,colorlinks=true,linkcolor=blue,urlcolor=blue,citecolor=blue]{hyperref}
\usepackage[sorting=none,style=numeric,doi=false,isbn=false,url=false,eprint=false]{biblatex}
\renewbibmacro{in:}{}
\usepackage[utf8]{inputenc}
\DeclareUnicodeCharacter{2215}{-}
\DeclareUnicodeCharacter{0327}{-}

\begin{document}

\title{Quantum teleportation by utilizing helical spin chains for sharing entanglement}

\author{Harshit Verma        \and
         Levan Chotorlishvili \and  Jamal Berakdar \and Sunil Kumar Mishra 
}

\authorrunning{H. Verma        \and
         L. Chotorlishvili \and  J. Berakdar \and Sunil K. Mishra} 

\institute{H.Verma \at Centre for Engineered Quantum Systems (EQUS), School of Mathematics and Physics, The University of Queensland, St Lucia, QLD 4072, Australia\\
              \email{h.verma@uq.edu.au}
           \and
            L. Chotorlishvili \at Institut f\"ur Physik, Martin-Luther-Universit\"at Halle-Wittenberg, D-06099 Halle, Germany
             \and
             J. Berakdar \at  Institut f\"ur Physik, Martin-Luther-Universit\"at Halle-Wittenberg, D-06099 Halle, Germany
             \and
             Sunil K. Mishra \at Department of Physics, Indian Institute of Technology (Banaras Hindu University), Varanasi - 221005, India\\
             \email{sunilkm.app@iitbhu.ac.in}
}

\date{Received: date / Accepted: date}

\maketitle

\begin{abstract}
We develop a new protocol for sharing entanglement (one ebit) between two parties using the natural dynamics of helical multiferroic spin chains. We introduce a novel kicking scheme of the electric field for enhancing the teleportation fidelity in our protocol that works in the presence of an appropriate choice of parameters. We also investigate the effect of a common spin environment causing decoherence in the entanglement sharing channel. We compare the results to that of XXZ and XX models subject to a similar entanglement sharing protocol and find that the helical multiferroic chain with the kicking scheme provides a better singlet fraction. We show that the kicking scheme in conjugation with the optimized parameters enhances the fidelity of teleportation even in the presence of impurities and/or decoherence. The advantage of the kicking scheme shown in the impurity cases is an important result to be useful in a realizable setup of helical multiferroic spin chain.
\end{abstract}

\section{Introduction}
The basic question of transmitting quantum information has been central
to the development of various quantum communication protocols
such as quantum teleportation \cite{benn}, direct state
transfer \cite{bose} and others. A variety of systems have been
studied as quantum channels which enable the transfer of qubits
using the often-cited
protocols \cite{Wilde,Adesso,Ivan,Rigolin,Greplova,Fortes,Campos,venuti,venutixx}. Many
types of spin chain systems have been found to be effective
for both teleportation and direct transfer
\cite{Albanese2004,Christandl2005,Boness2006,Banchi2010,Apollaro2012,tele_arxiv}. A related development has been in routing quantum information through network of spin chains \cite{rout1,rout2,rout3,rout4,Hu2009} while studying the effect of phase of the spin chain, quench, and external field. 

On the experimental side, with the  remarkable progress in nanotechnology and material science during the last two decades, several quantum information protocols have become experimentally realizable in spin chain systems. For instance, solid-state-based one-dimensional and quasi-one-dimensional multiferroic spin chains (such as $\rm LiCu_{2}O_{2}$) were experimentally realized. In the experimentally investigated system by Menzel {\it et.~al.} \cite{Menzel2012}, it has been shown (using spin-polarized  scanning tunneling spectroscopy), that  the effect of changing  the spin direction at one site at the edge of the  chain can be accessed  on the other  end of the chain. 
A key feature of the particular spin chain system studied here is that it possesses an intrinsic electric polarization associated with  spin non-collinearity. This ferroelectric property of the quantum quasi-one-dimensional $S=1/2$ magnet such as  $LiCu_{2}O_{2}$ was  experimentally verified  \cite{Park}, and allows us to act on the chain with an external electric field. In addition to  $LiCu_{2}O_{2}$,  there are a number of other  helical multiferroics  materials  with  intrinsic coupled magnetic and ferroelectric order parameters \cite{Mostovoy,Park,Nagaosa,Chotorlishvili,Sekania,s5,s6,s7} which were also  shown to be useful in quantum information processing \cite{ol,hv}, among various other applications.

The focus of this manuscript is on addressing the question that whether it is possible to achieve high quality teleportation using helical spin chains as an entanglement sharing channel? We have also studied that whether the magneto-electric coupling associated with helical multiferroics can be functionalized  for improving the fidelity of quantum teleportation.

In a realistic setting, the entanglement sharing channel such as a spin chain may have
embedded impurities and may also be susceptible to noise and other environmental effects. These factors are expected to affect the transfer of quantum
information and therefore, would influence the fidelity of information transfer.
Specifically, in quantum teleportation, noise may set
in at any point of time which may lead to the conversion of a pure entangled state (to be
shared initially) into a mixed state or inaccurate
detection of the shared entangled state by sender and/or receiver \cite{Oh2002}. In this regard, local environment has been seen to have effects that are sometimes
counter-intuitive such as an increase in the teleportation fidelity \cite{horo,bando,Yeo_2005,YY,Yeo_2009,ishi}. 

In this manuscript, we have proposed a generic protocol wherein the entanglement is generated and encoded in the middle of a spin chain and it becomes available at the ends through the intrinsic dynamics of the chain. The parties involved in teleportation have access to the ends of the chain and hence, the available entanglement is utilized by the sender and receiver for the teleportation task. Here we show that helical multiferroics can indeed be used to share the entanglement such that an appreciable fidelity is achieved. In general, we have considered systems with helical spin order and identified the system parameters required for high-fidelity teleportation. Moreover using a Floquet map, we have studied the system dynamics numerically in the presence of kicked electric field which has been found to increase the teleportation fidelity if an appropriate set of parameters is chosen. Additionally, we have considered the effect of impurities and uniform environment on the teleportation fidelity.

The paper is organized in the following manner: at first we discuss
the spin chain system with helical multiferroic Hamiltonian and its characteristics,
particularly the various interaction terms. We also present the unique kicking
scheme that we would use throughout the manuscript. Afterwards, we discuss the quantum
teleportation protocol highlighting the role of the entanglement as a resource and
addressing the question of how does the quality of teleportation depend on this
resource. The formula for teleportation fidelity -- a measure that quantifies
the quality of teleportation is derived for our specific protocol and also
in the setting of an environment which causes decoherence. In the next
section, we discuss the numerical results with kicked electric field to demonstrate
that a higher fidelity can be achieved using this scheme. In this respect,
we also study the cases with specific types of impurities
as earmarked in \cite{hv} and a specific type of environment
which causes decoherence in the system \cite{cuch,Cai,Hu2009}.  Moreover, we compare 
the results obtained to that of XX and XXZ models found in previous studies as well
as when subject to our entanglement sharing protocol.
\section{Models}
We study a one-dimensional multiferroic chain (along the x-axis) of localized 
spins $\vec{S}_{i}$ modelled by a $J_1$--$J_2$ Hamiltonian and also including the interaction 
energy of an external electric field $\mathcal{E}$ coupled to the spin-driven electric
polarization of the chain. The Hamiltonian of this chain is given by:
\begin{eqnarray}
\mathcal{H}&=&-J_{1}\sum_{i}\vec{S}_{i}.\vec{S}_{i+1}-J_{2}\sum_{i}\vec{S}_{i}.\vec{S}_{i+2}+\mathcal{E}(t)g_{ME}\sum_{i}{(\vec{S}_{i}\times \vec{S}_{i+1})}^z.
\label{eq:hamil}
\end{eqnarray}
The first two terms stand for the Heisenberg type exchange interactions of the 
spin $\vec{S}_{i}$ with nearest ($\vec{S}_{i+1}$) and next nearest neighbors ($\vec{S}_{i+2}$) with different coupling constants i.e. $J_1$ and $J_2$ respectively. Taking $J_1 > 0 $ i.e. ferromagnetic nearest neighbor interaction and $J_2 <0$, i.e. anti-ferromagnetic next nearest interaction leads to frustration and helical spin order. The third term is the coupling (with a magnetoelectric coupling constant $g_{ME}$) of an electric field $\mathcal{E}$ applied along the $y$ direction to the spin-driven ferroelectric polarization. We note that the ferroelectric polarization of one-dimensional chiral multiferroic chains can be easily controlled and switched in an experiment \cite{Schrettle2008}.

\begin{figure}
\begin{minipage}[c][9cm][c]{0.5\columnwidth}
\centering
\includegraphics*[trim =9.5cm 6cm 13.5cm 6.5cm, clip, scale=0.55]{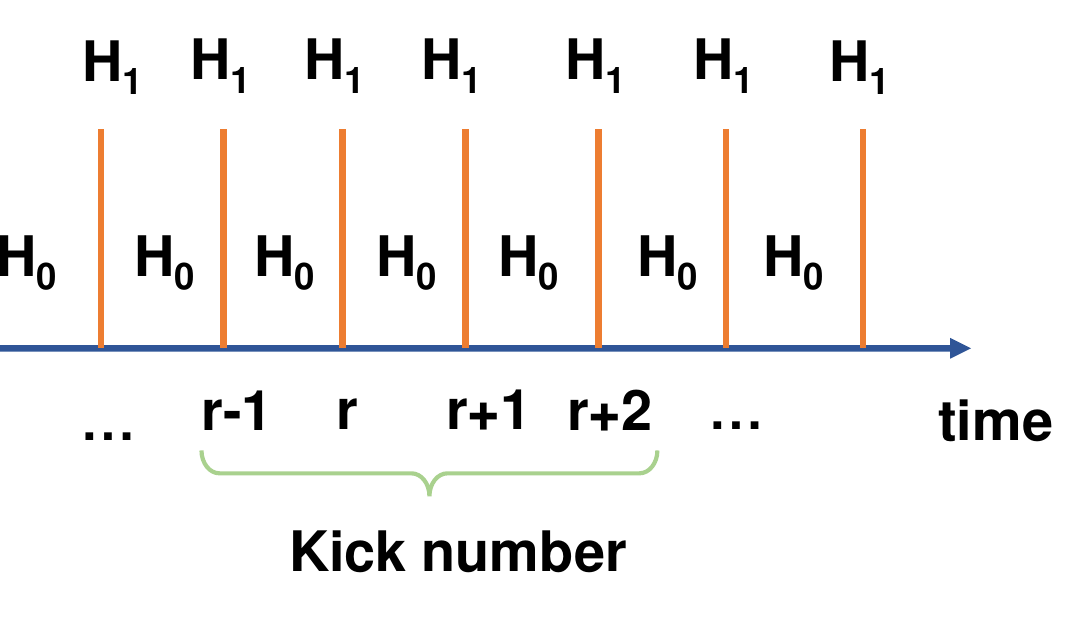}
\subcaption{The train of pulses and the notation of the kicking scheme used in the manuscript.}
\label{kick}
\addtocounter{subfigure}{1}
\includegraphics*[trim =4.5cm 10cm 6cm 5cm, scale=0.32]{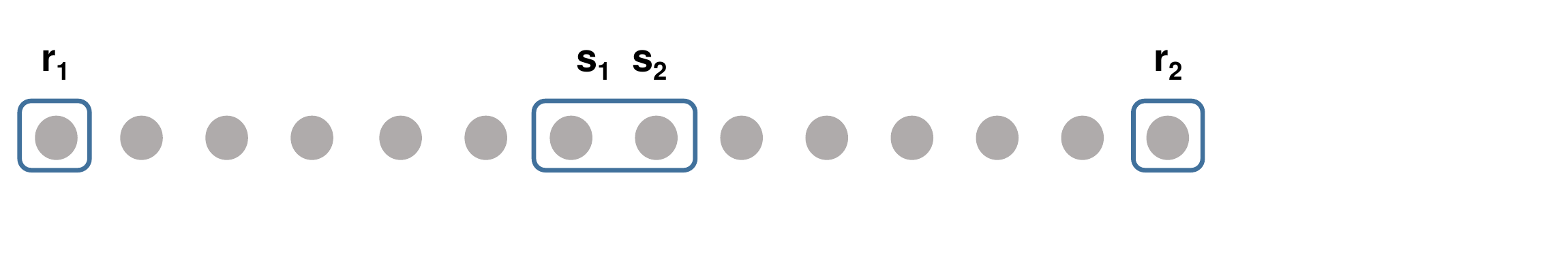}
\subcaption{Schematic showing the spin chain system with the sender sites marked by s$_1$, s$_2$ and receiver sites by r$_1$, r$_2$. A Bell pair is generated at s$_1$, s$_2$ and we expect to use the entanglement available at r$_1$, r$_2$ at a later time for quantum teleportation.}
\label{fig:process13}
\end{minipage}
\hspace{0.25cm}
\begin{minipage}[c][9cm][t]{0.5\columnwidth}
\centering
\includegraphics*[trim =9.5cm 5cm 9.5cm 2cm, scale=0.4]{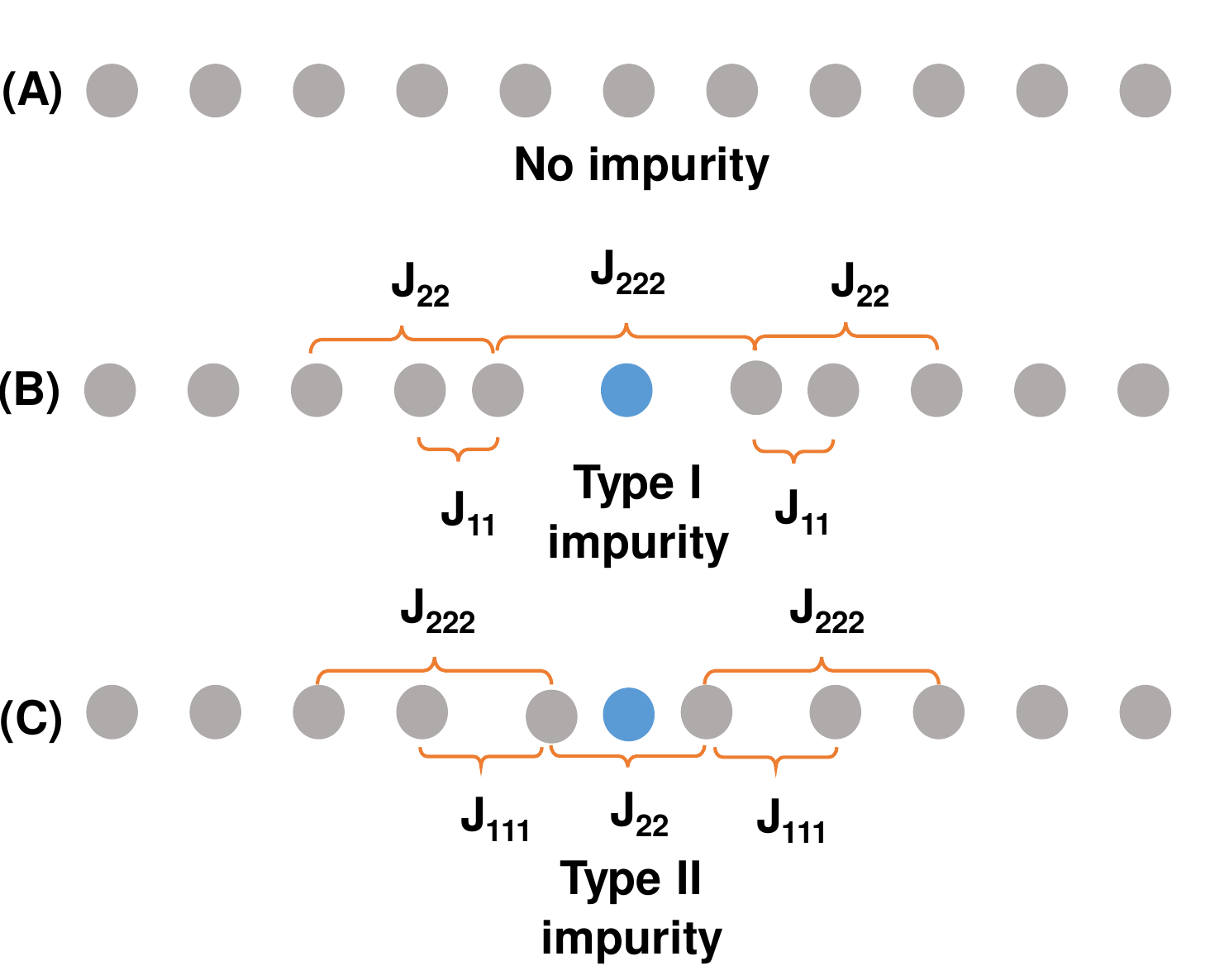}
\addtocounter{subfigure}{-2}
\subcaption{Schematic showing the effect of introduction of Type I and Type II impurities in the spin chain. The interactions which are marked ($J_{11}/J_{22}/J_{111}/J_{222}$) undergo deviation from normal interactions which are $J_1$ and $J_2$. (A) shows a normal spin chain with no embedded impurity. (B) shows a spin chain with Type I impurity embedded in its middle. (C) shows a spin chain with Type II impurity embedded in its middle.}
\label{fig:imp}
\end{minipage}
\caption{}
\label{fig0}
\end{figure}

As seen from Eq.~\ref{eq:hamil}, helical multiferroic systems possess two coupled order parameters: magnetization and ferroelectric polarization. The latter can be acted upon by an external field and hence, the spin dynamics can be steered by means of an electric field. The key element in this regard is the $z$ component of the vector chirality $(\vec{S}_{i}\times \vec{S}_{i+1} )^z$.
We study the following cases: (1) a static electric field ($\mathcal{E}=\mathcal{E}_0$), and (2)
apart from $\mathcal{E}_0$ we apply a train of kicks with a period $\tau$ and amplitude $\mathcal{E}_1$ along $y$-direction. The underlying assumption
in the second case is that the duration of one pulse in the train is much shorter
than the intrinsic timescale of the system (set by $J_1$, $J_2$, $g_{ME}$ and is
typically in the picosecond time range) and thus, our pulse train has a structure shown
in Fig.~\ref{kick}. Such a train of electric filed can be achieved in effect by highly assymmteric single-cycle THz electric pulses having an appropriate profile providing for a sharp pulse for a short time and a static-dc electric field at later times \cite{Sekania, kickrev}. Henceforth, all occurrences of time would be considered in the units of $1/J_1$. 

Under the kicked field, the Hamiltonian can be split into static and dynamic parts as follows:
\begin{eqnarray}
\mathcal{H}&=&\mathcal{H}_0+\mathcal{H}_1 (t),\nonumber\\
\mathcal{H}_0&=&-J_{1}\sum_{i=1}^{N-1}\vec{S}_{i}.\vec{S}_{i+1}-J_{2}\sum_{i=1}^{N-2}\vec{S}_{i}.\vec{S}_{i+2}+ E_0\sum_{i=1}^{N-1}{(\vec{S}_{i}\times \vec{S}_{i+1})}^z,\\
\mathcal{H}_1&=&E_1\sum_{n=1}^{n=n_{max}} \delta \big(t- n \tau \big)\sum_{i=1}^{N-1}
{(\vec{S}_{i}\times \vec{S}_{i+1})}^z.
\end{eqnarray}
Here, $N$ is the number of sites in the chain considered, $n$ indexes the kicks, $n_\textrm{max}$ is the number of kicks, $E_0=g_{ME}\mathcal{E}_0$, and $E_1=g_{ME}\mathcal{E}_1$. The open boundary condition of the spin chain is apparent from the range of index $i$ for the different terms of the Hamiltonian.
Starting at $t=0$ from an initial state, say $~|\psi(t=0)\rangle$ the time evolution operators:
\begin{eqnarray}
\hat{\mathcal{U}_0}&=&\exp\bigg(iJ_{1}\tau\sum_{i=1}^{N-1}\vec{S}_{i}.\vec{S}_{i+1}+iJ_{2}\tau\sum_{i=1}^{N-2}\vec{S}_{i}.\vec{S}_{i+2} \nonumber-iE_0\tau\sum_{i=1}^{N-1}{(\vec{S}_{i}\times \vec{S}_{i+1})}^z \bigg),
\nonumber \\
\hat{\mathcal{U}}_1&=&\exp\bigg(-iE_1 \sum_{i=1}^{N-1}{(\vec{S}_{i}\times \vec{S}_{i+1})}^z\bigg),
\label{eq:impulse}
\end{eqnarray}
 deliver the state just after the $r^{th}$ kick (or at time $t=r\tau$) as
\begin{eqnarray}
 |\psi(t=r\tau)\rangle=\big(\hat{\mathcal{U}}_1~\hat{\mathcal{U}}_0\big)^r~|\psi(t=0)\rangle .
\end{eqnarray}
In comparison to the multiferroic Hamiltonian given in Eq.~\ref{eq:hamil}, the Hamiltonian of XXZ model has the following generic form:
\begin{equation}
H= J_x \sum_{i=1}^{N-1} \big(\hat{S}_i^x\hat{S}_{i+1}^{x} +\hat{S}_i^y\hat{S}_{i+1}^{y}\big) +J_z\sum_{i=1}^{N-1} \hat{S}_i^z\hat{S}_{i+1}^{z}~,
\label{eq:hamXXZ}
\end{equation}
where $J_x$ and $J_z$ are the nearest neighbour exchange interactions. The Hamiltonian in Eq.~\ref{eq:hamXXZ} reduces to that of the XX model when $J_z = 0$.

We subscribe to the impurity models discussed in Ref. \cite{hv} and borrow the effects of specific impurities considered therein as well as the terminology. Fig.~\ref{fig:imp}, demonstrates the effects of Type I and Type II impurities set amidst a spin chain. For  both the types of impurities, an assumption is made -- the nearest and next nearest neighbour interactions involving the impurity do not change. Also, the magnetoelectric coupling is assumed to remain the same. The effects of impurity are confined to $3$ sites near the impurity as seen from Fig.~\ref{fig:imp}. The impurities cause elongation and contraction of various ``bonds" resulting in corresponding changes in the interaction strengths $J_1$ and $J_2$.
The term $J_{22}/J_{2}$ considered in all of the figures concerning the effect of impurities is indicative of
the impurity strength and refers to a particular bond near the embedded impurity
which can be ascertained from Fig.~\ref{fig:imp}. For the case of Type I impurity, increment in $J_{22}$ also causes similar increment in $J_{11}$ and decrement in $J_{222}$. Similarly, in case of Type II, increment in $J_{22}$ happens simultaneously with decrement in $J_{111}$ and $J_{222}$. We have assumed that the factor of increment of $J_{22}$ in all the cases is the same as that of the factor of change in $J_{11}$, $J_{222}$ and $J_{111}$.

\section{Quantum Teleportation Fidelity}
\begin{figure}
\centering
\includegraphics*[trim =3cm 1cm 8cm 2cm, scale=0.35]{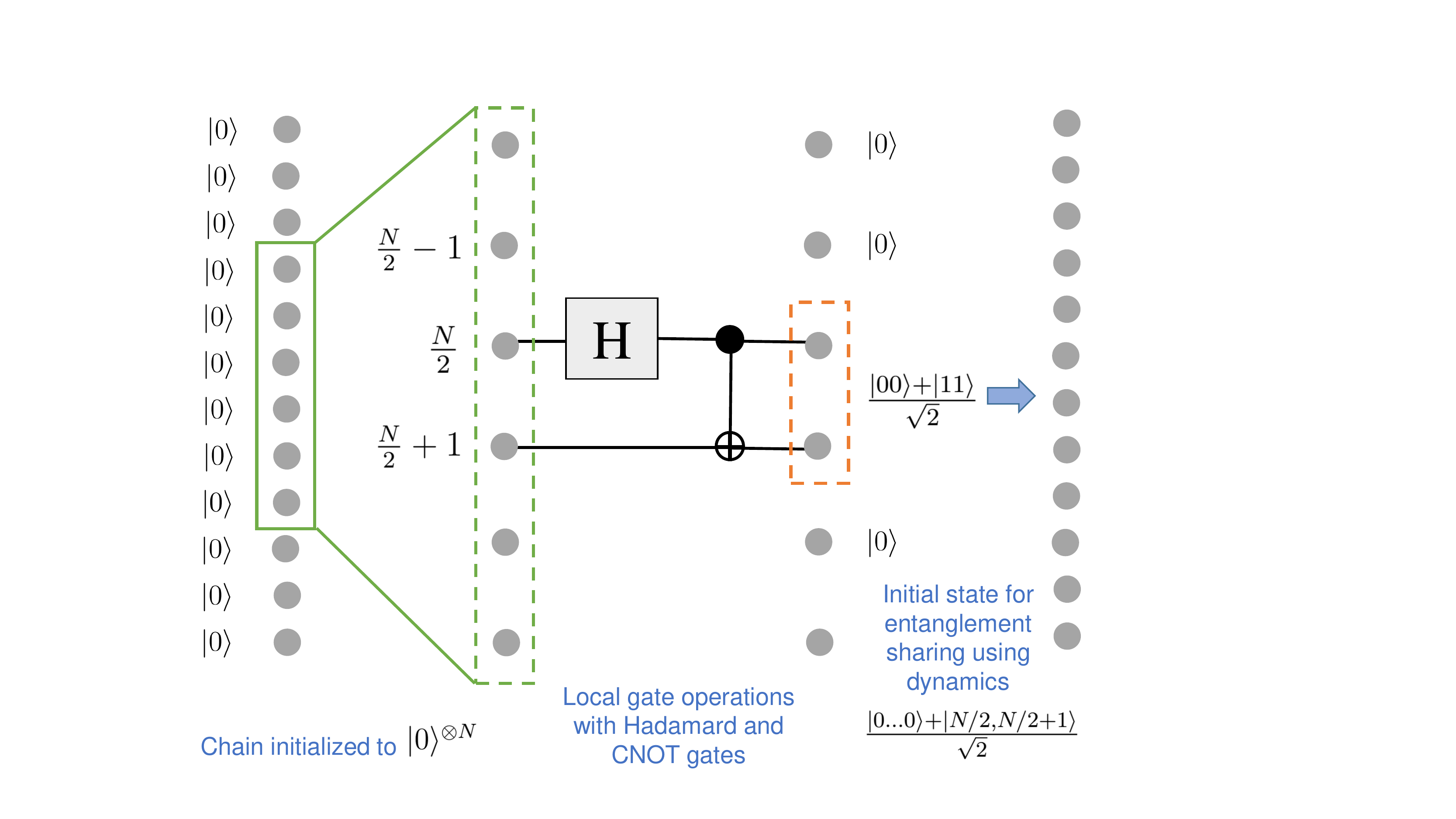}
 \caption{The process of using local gates in the middle of the initialized spin chain to prepare a Bell pair which then propagates.}
 \label{gate}
\end{figure}

Quantum teleportation is a well-known protocol for transferring quantum information between two parties \cite{benn}, which can be accomplished by using an entangled pair of qubits, a quantum channel and two bits of classical information sent via a classical channel. This teleportation scheme is generally referred to as \textit{standard teleportation scheme (STS)}. One of the key steps involved in quantum teleportation of a single qubit is the sharing of one qubit each of an entangled pair of qubits, preferably a Bell-pair, between the sender and the receiver. The quality of teleportation is quantified by the fidelity between the intended qubit and the qubit received. This fidelity is dependent on the singlet fraction which is defined in \cite{Horodecki,Li2013} as follows:
\begin{equation}
f = \langle \Omega^{00}\vert \chi\vert\Omega^{00}\rangle,
\end{equation}
where $\chi$ denotes the state of the qubits carrying
entanglement (here, the end qubits -- $r_1$, $r_2$ in Fig. \ref{fig:process13}) and $|\Omega^{00}\rangle = \frac{\vert 00\rangle + \vert 11\rangle}{\sqrt{2}}$ is
a Bell pair. In principle, the maximum singlet fraction can be obtained by considering all the possible Bell pairs in the above equation rather than just $|\Omega^{00}\rangle $.
The fidelity of teleportation, as discussed above, has been found to be related to singlet fraction in the following way \cite{Horodecki,Li2013} :
\begin{equation}
F=\frac{2 f+1}{3}.
\label{f,sf}
\end{equation}
In our protocol, we assume that initially, a Bell pair ($|\Omega^{00}\rangle$) is generated and encoded at the middle of
the spin chain. This can be achieved, in principle by using local gates -- particularly Hadamard and CNOT gates in succession on the middle qubits (shown in Fig. \ref{gate}).

Through the natural dynamics of the spin chain, it is to be expected that significant entanglement will be
available at the ends of the spin chain with the sender and receiver at a later time, which they could utilize in teleportation. This proposition is motivated by the prospect of a commercial arrangement wherein the ``service provider" has a mechanism to provide entanglement (or ideally one ebit of entanglement) to each party, thus, controlling the quantum teleportation and hence, being able to commercialize its ``service".
Various other systems such as XX spin chains and anti-ferromagnetic spin chains \cite{Campos,Giampaolo,tele_arxiv} have been shown to be effective for the task of availing entanglement preceding teleportation. However, the mechanisms used  to generate and distribute entanglement in the aforementioned references are different than our case. For example, in \cite{Campos} a dimerized chain is used which possesses ground states having singlet pairs on alternate pairs of sites. By changing a set of parameters, the ground state of the system is made to possess singlet pairs with a `global' singlet between first and the last sites. Noisy quantum channels have also been studied in this regard and have been shown to be effective \cite{Oh2002}.

Let us consider the case with an even number of sites in a spin chain. We label the chain sites starting from the left as $1,2,\dots,N/2,N/2+1,\dots,N-1,N$.   We assume that
the chain is initially prepared in all
spin up state i.e. $\vert 0 0 \dots 0 0\rangle$ and a Bell pair
is generated and encoded at sites $N/2$ and $N/2+1$ (see Fig.~\ref{fig0}). This initial state can be constructed by applying a Hadamard gate on the $N/2$ site followed by a CNOT gate with $N/2$ site as controlled site and $N/2+1$ site as target site (see Fig.~\ref{gate}). In experiment this can be constructed by applying a strong constant magnetic field to align all spins in up direction, and followed by applying rf field  $\pi$ pulse to rotate the target spin at $N/2+1$ site to generate a Bell state $\vert\Omega^{00}\rangle$ in the middle \cite{Zajac439}. At $t>0$ we switch off the strong constant magnetic field and let the spins evolve through their internal dynamics.
 The state of the system at $t=0$ can be written as follows:
\begin{eqnarray}
\vert \psi(t=0)\rangle &=& \vert 0 \dots 0\rangle \otimes \vert\Omega^{00}\rangle \otimes \vert 0 \dots0\rangle =\frac{\vert 0 \dots 0 \rangle + \vert N/2, N/2 +1\rangle}{\sqrt{2}},
\end{eqnarray}
where $\vert N/2, N/2+1 \rangle $ refers to spin flipped state from $\vert 0\rangle $ to $\vert 1\rangle$ at sites $N/2$ and $N/2+1$.
Note that the Hamiltonian in Eq.~\ref{eq:hamil} commutes with the total magnetization, i.e. $[M,H]=0$. The total magnetization is given through the total $Z$ component of spins at all sites which is $M=\sum\limits_{i=1}^N\hat{S}^{z}_{i}$. Therefore, the time evolution of such a system does not mix the different spin (magnetization) sectors. Hence, after time evolving $\vert\psi (t=0)\rangle$ we obtain:
\begin{eqnarray}
\vert \psi(t)\rangle &=& e^{\frac{-i H t}{\hbar}} \vert\psi(0)\rangle=\frac{G|0\rangle +G|N/2, N/2+1\rangle}{\sqrt{2}},
\label{eq10}
\end{eqnarray}
where $G \equiv \exp{\frac{-i H t}{\hbar}} $ is the time evolution operator. Because of the commutation of $M$ with $H$, we know that
\begin{eqnarray}
G|0\rangle \rightarrow e^{-iE^0t}|0\rangle, \qquad G|N/2, N/2+1\rangle \rightarrow |j, j'\rangle,
\label{timeevol}
\end{eqnarray}
where $\vert j, j' \rangle $ refers to a general
spin flipped state from $\vert 0,0\rangle $ at sites $j$ and $j'$ only, meaning
a general state with 2 spins down and other spins up, and $E^0$ is the eigenenergy of the state $|0\rangle$ with respect to the Hamiltonian $H_0$ (Eq.~\eqref{eq:hamil}).
Therefore, the time evolved state can be expressed as:
\begin{eqnarray}
\vert \psi(t)\rangle = \frac{e^{-iE^0t}\vert 0 \rangle }{\sqrt{2}} + \frac{1}{\sqrt{2}}\sum_{j}\sum_{j'}\langle j, j'\vert G \vert N/2, N/2+1 \rangle \vert j, j'\rangle.
\end{eqnarray}
It is noteworthy that the extra phase in the $|0\rangle$ state will be the same irrespective of the kicked or unkicked time evolution used, once the spin chain parameters are fixed.
\begin{figure*}[t]
  \begin{subfigure}{0.5\textwidth}
    \includegraphics*[trim =4.5cm 7cm 10cm 5.5cm, width=\textwidth]{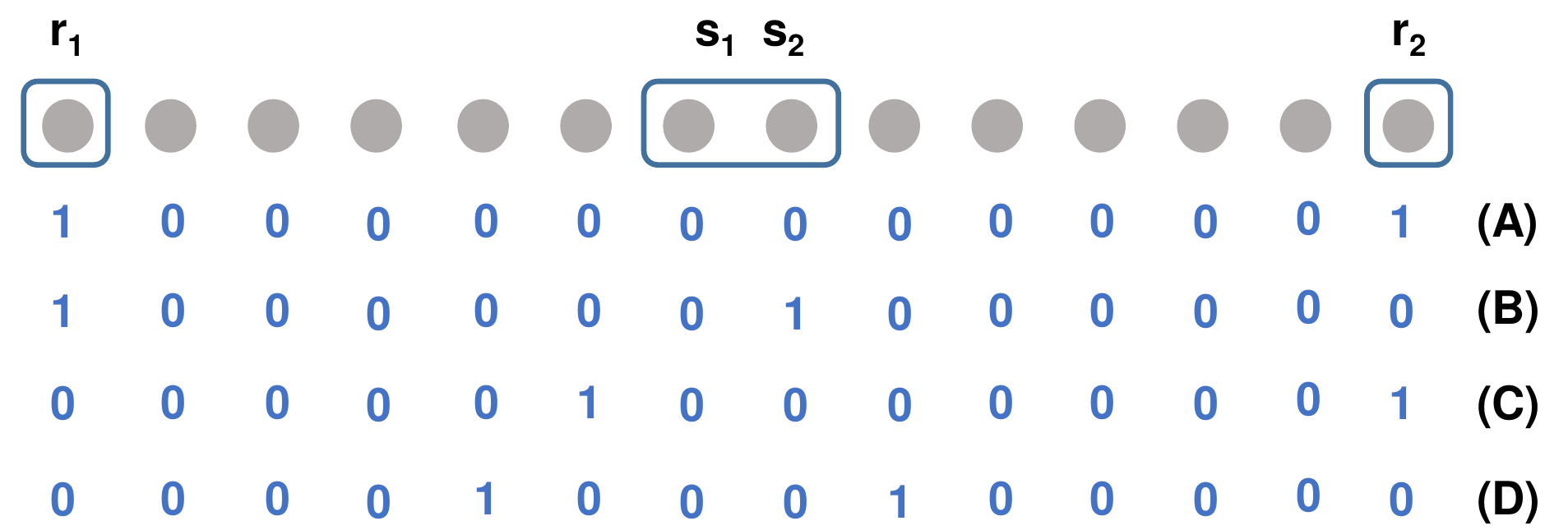}
    \caption{Partial trace to obtain the density matrix of the edge qubits ($\rho_\textrm{out}$) results in the following possibilities: (A) type 1: Partial trace yields $|11\rangle $ -- a single possible case. (B) type 2: $|10\rangle $ resulting from a partial trace. N-2 cases of such type are possible. (C) type 3: Partial trace gives $| 01 \rangle $. N-2 cases of such a type possible (D) type 4: $|00\rangle$ obtained after partial trace. $ ^{N-2} C_2$ possible cases. So in total, $1+(N-2)+(N-2)+^{N-2}C_2$ cases which adds up to $^NC_2$ as expected.}
    \label{type}
  \end{subfigure}
  \hspace{0.25 cm}
  \begin{subfigure}{0.5\textwidth}
    \includegraphics*[trim =2.5cm 8cm 9cm 7cm, width=\textwidth]{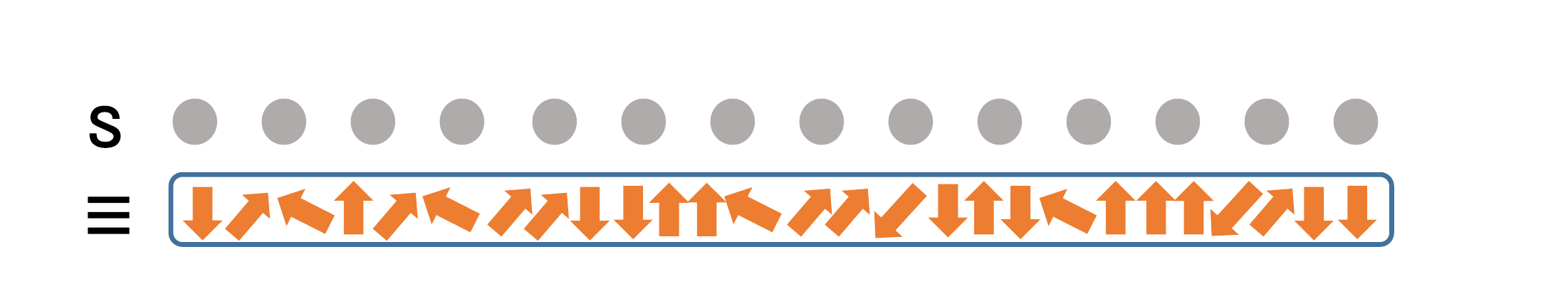}
    \caption{S denotes the main system consisting of a helical spin chain that
interacts with the environment $\Xi$. Note that the number of spins in the environment is more than that in the main system}. The coupling $g_k$ determines
the magnitude of interaction between the system and the
environment. We consider a uniform environment with the same
coupling constant at various sites of the environment. The Hamiltonian of
the main system is $H_F$, that of the system interacting with
environment is $H_{S\Xi}$ and of the complete system is $H_{total}$.
    \label{env}
  \end{subfigure}
  \caption{}
\end{figure*}

The corresponding output density matrix i.e.\ for the receiver qubits at both the ends is obtained by tracing over all other qubits and is given as follows:
\begin{eqnarray}
\rho_{out} &=& Tr_{2,3,\dots,N-2,N-1} \big[ \rho (t)\big] \nonumber\\
&=& \frac{1}{2}\vert 00 \rangle\langle 00\vert+ \frac{1}{2}\sum_{j, j' (type 4)}\vert\langle j, j' \vert G \vert N/2, N/2 +1\rangle\vert ^2 |00\rangle\langle 00|\nonumber\\
&+& \frac{1}{2}\sum_{j, j' (type 3)}\vert\langle j, j' \vert G \vert N/2, N/2 +1\rangle\vert ^2 |01\rangle\langle 01| \nonumber\\
&+& \frac{1}{2}\sum_{j, j' (type 2)}\vert\langle j, j' \vert G \vert N/2, N/2 +1\rangle\vert ^2 |10\rangle\langle 10| \nonumber \\
&+& \frac{1}{2}|\langle 1,N |G| N/2, N/2 +1 \rangle |^2|11\rangle\langle 11|\nonumber\\
&+& \frac{1}{2}e^{-iE^0t}\langle 1,N |G| N/2, N/2 +1 \rangle ^*|00\rangle\langle 11| \nonumber \\
&+& \frac{1}{2}e^{iE^0t}\langle 1,N |G| N/2, N/2 +1 \rangle |11\rangle\langle 00|,
\end{eqnarray}
where, type 2 refers to all possible spin chains with the first
qubit definitely as $| 1 \rangle $ and any one of other
qubits except for the last as $|1\rangle $ which has N-2 possibilities
in total and would yield $|10\rangle $ after partial tracing. Type 3 refers to all
possible chains with the last qubit definitely $|1\rangle$ and anyone of
the other qubits except for the first as $|1\rangle$ which has N-2 possibilities
in total and yields $|01\rangle $ after partial tracing. Type 4 refers to all possible
spin chains with neither of first or last qubits ($r_1$ and $r_2$) as $|1\rangle$ but any two of
other qubits as $|1\rangle$ which has a total of $^{N-2}C_2$ possibilities
yielding $|00\rangle$ on partial tracing. Fig.~\ref{type} gives a pictorial representation of all cases and indicates the number of possibilities.

Finally, for $|\psi_{in}\rangle = |\Omega^{00}\rangle$, the singlet fraction (f) is given as:
\begin{eqnarray}
f &=& \langle \psi_\textrm{in}|\rho_\textrm{out}| \psi_\textrm{in} \rangle \nonumber \\
&=& \frac{1}{4} + \frac{1}{4}\sum_{j, j' (type 4)}\vert\langle j, j' \vert G \vert N/2, N/2 +1\rangle\vert ^2 + \frac{1}{4}|\langle 1,N |G| N/2, N/2 +1 \rangle |^2 \nonumber \\
&+& \frac{1}{2} Re \left[ e^{iE^0t} \langle 1,N |G| N/2, N/2 +1 \rangle\right].
\label{sinfrac}
\end{eqnarray}
\section{Singlet fraction with uniform spin environment}
We will now introduce a realistic setting where a uniform spin environment
acts on the multiferroic spin chain system, thereby altering its dynamics.
Here we are considering a uniform spin environment which is formed by a spin chain of length $P$ (having P sites) \cite{cuch,Cai,Hu2009}. 

We can consider the spin chain is placed on the substrate of a long one dimensional spin chain of P spins.  
In this arrangement it is easy to visualise the coupling of our chain with the quantum substrate (see Fig.~\ref{env}).  Denoting the Hamiltonian of the main system by $H_F$, the part of Hamiltonian for the coupling between the environment and the main system is
\begin{equation}
H_{S\Xi} =  \underbracket{2\sum^N _{i=1} \hat{S}^z_i}_{main~system} \otimes \underbracket{\sum^P_{k=1} g_k~ \hat{S}_k^z }_{environment}.
\label{inte}
\end{equation}
This whole system, including the main system and the environment is
represented in Fig.~\ref{env}. The Hamiltonian of the complete
system is given by
\begin{equation}
H_\textrm{total}= H_F + H_{S\Xi} .
\label{totalham}
\end{equation}
Here, we have assumed that the environment is not self-interacting. This model of the
environment and the particular model of interaction with the main system has been studied
before in the case of a general quantum information transfer channel \cite{Cai}. The noise is based on a “central spin interaction” model introduced in \cite{cuch}. Owing to its adoption in the analysis of other schemes concerning quantum information transfer protocols, and the fact that its effect in our entanglement sharing scheme can be analytically calculated is one of the reasons that we considered this particular model. Moreover, the physical basis of the environment model is rooted in the decoherence noise experienced by spin chains and/or many body physical platforms viz. coupled quantum dots, under experimental conditions.

For the derivation of singlet fraction in a system where the middle two
spins are substituted by a Bell pair and the system interacts with the environment
as given by Eq.~\ref{inte}, we assume that the environment is initially
in the state $\vert \psi_{\Xi} (0)\rangle=\sum^{2^P-1}_{m=0} c_m \vert m \rangle$. Here, $\vert m \rangle$ represents
the spin basis of an environment consisting of P sites. Therefore, in this case, the output density
matrix is given as
\begin{equation}
\rho_\textrm{out} = Tr_{2,3,\dots,N-2,N-1}\bigg[Tr_{\Xi}\big(\rho(t)\big)\bigg] ,
\end{equation}
where $\rho(t)$ is the density matrix of the complete system at a general time $t$. It has been assumed that initially, the system and the environment are in a non-entangled state.
So, following the steps in the previous derivation, we have, at time $t=0$,
\begin{eqnarray}
\vert \psi_\textrm{total} (0)\rangle &=& \vert \psi_S (0)\rangle \otimes \vert \psi_{\Xi} (0)\rangle=\bigg[ \frac{|0\rangle + | N/2,N/2+1\rangle}{\sqrt{2}}\bigg] \otimes \sum^{2^P-1}_{m=0} c_m \vert m \rangle ,\nonumber\\
\end{eqnarray}
where $\vert \psi_\textrm{total}\rangle$ represents the state of the full system consisting of the main system (with its state represented by $\vert \psi_S \rangle$) and the environment (with its state represented by $\vert \psi_{\Xi} \rangle$)
Therefore, at a later time $t$,
\begin{eqnarray}
\vert \psi_\textrm{total}(t)\rangle &=& \frac{G \vert 0 \rangle}{\sqrt{2}}\vert \epsilon_0 (t)\rangle + \frac{\sum_{j}\sum_{j'}\langle j, j'\vert G \vert N/2, N/2+1 \rangle \vert j, j'\rangle}{\sqrt{2}} \vert \epsilon_{j,j'} (t) \rangle , \nonumber \\
\end{eqnarray}
where G is the time evolution operator of the main system which has been previously indicated in Eq.~\ref{timeevol}. Also, $\vert \epsilon_0 (t) \rangle $ and $\vert \epsilon_{j,j'} (t) \rangle $ are defined below.
\begin{eqnarray}
\vert \epsilon_0 (t)\rangle = \sum^{2^P-1}_{m=0} c_m e^{-iNtB_m} \vert m \rangle, \qquad
\vert \epsilon_{j,j'} (t) \rangle= \sum^{2^P-1}_{m=0} c_m e^{-i(N-2)tB_m} \vert m \rangle,
\label{epsil}
\end{eqnarray}
which have been calculated using the action of $H_{S\Xi}$ on $|\psi_{\Xi} (t)\rangle$. This depends specifically on the state of the environment. For the environment in an eigenstate $\vert m \rangle = \vert m_1, m_2, ..., m_P\rangle$, $\sum_{k=1}^P g_k \sigma_k^z \vert m\rangle$ gives $B_m$ as follows:
\begin{equation}
B_m= \sum^P_{k=1} \frac{1}{2}(-1)^{m_k} g_k .
\label{bm}
\end{equation}
The density matrix of the system obtained after tracing over the environment is as follows:
\begin{eqnarray}
Tr_\Xi\big[\rho (t)\big] &=& \frac{1}{2} e^{-iE^0t} \vert 00 \rangle\langle 00 \vert+ \frac{1}{2} \sum_j \sum_{j'} |\langle j,j' \vert G\vert N/2, N/2+1\rangle |^2 \vert j,j'\rangle\langle j,j'\vert \nonumber \\
&+& \frac{1}{2} e^{-iE^0t} \sum_j \sum_{j'} \langle j , j' \vert G \vert N/2, N/2+1\rangle ^* \vert 0 \rangle\langle j,j'\vert r^* (t) \nonumber \\
&+& \frac{1}{2} e^{iE^0t}\sum_j \sum_{j'} \langle j , j' \vert G \vert N/2, N/2+1\rangle \vert j,j' \rangle\langle 0\vert r(t) ,
\label{rhoet}
\end{eqnarray}
where,
\begin{eqnarray}
r(t) &=& \langle \epsilon_0 (t)\vert \epsilon_{j,j'}(t)\rangle= \sum^{2^P-1}_{m=0} |c_m|^2 e^{-i(N-2)tB_m} e^{iNtB_m} \equiv \sum^{2^P-1}_{m=0} |c_m|^2 e^{2itB_m} .
\label{rt}
\end{eqnarray}
Finally, the output density matrix is obtained below.
\begin{eqnarray}
\rho_\textrm{out} &=& Tr_{2,3,\dots,N-2,N-1} \bigg[ Tr_\Xi\big[\rho (t)\big]\bigg] \nonumber\\
&=& \frac{1}{2}\vert 00 \rangle\langle 00\vert + \frac{1}{2}\sum_{j, j' (type 4)}\vert\langle j, j' \vert G \vert N/2, N/2 +1\rangle\vert ^2 |00\rangle\langle 00| \nonumber\\
&+& \frac{1}{2}\sum_{j, j' (type 3)}\vert\langle j, j' \vert G \vert N/2, N/2 +1\rangle\vert ^2 |01\rangle\langle 01| \nonumber \\
&+& \frac{1}{2}\sum_{j, j' (type 2)}\vert\langle j, j' \vert G \vert N/2, N/2 +1\rangle\vert ^2 |10\rangle\langle 10| \nonumber \\
&+& \frac{1}{2}|\langle 1,N |G| N/2, N/2 +1 \rangle |^2|11\rangle\langle 11| \nonumber \\
&+& \frac{1}{2} e^{-iE^0t}\langle 1,N |G| N/2, N/2 +1 \rangle ^*|00\rangle\langle 11| r^*(t) \nonumber \\
&+& \frac{1}{2} e^{iE^0t}\langle 1,N |G| N/2, N/2 +1 \rangle |11\rangle\langle 00| r(t) .
\label{rhoout}
\end{eqnarray}
Therefore, we obtain the singlet fraction as follows:
\begin{eqnarray}
f &=& \langle \psi_\textrm{in}|\rho_\textrm{out}| \psi_\textrm{in} \rangle \nonumber \\
&=& \frac{1}{4} + \frac{1}{4}\sum_{j, j' (type 4)}\vert\langle j, j' \vert G \vert N/2, N/2 +1\rangle\vert ^2 + \frac{1}{4}|\langle 1,N |G| N/2, N/2 +1 \rangle |^2 \nonumber \\
&+& \frac{1}{2} Re \left[ e^{iE^0t} \langle 1,N |G| N/2, N/2 +1 \rangle r(t) \right].
\label{fiden}
\end{eqnarray}
We shall consider a specific case wherein the environment is in
an all spin up state i.e. $\vert 0\dots 0 \rangle $. We also assume that
the coupling constant $g_k$ is, in fact, the same for all environment
sites referred by index $k$. We can now evaluate $r(t)$ with $c_0 =1$ and $B_m = \frac{P g}{2} $ where
$g$ is the uniform coupling constant considered. Therefore, $r(t) =e^{Pitg}$ can be
used in Eq.~\ref{fiden} to obtain the corresponding singlet fraction.

\section{Numerical results for kicked and unkicked multiferroic chains}
\begin{figure*}[t]
\includegraphics*[width=1\linewidth]{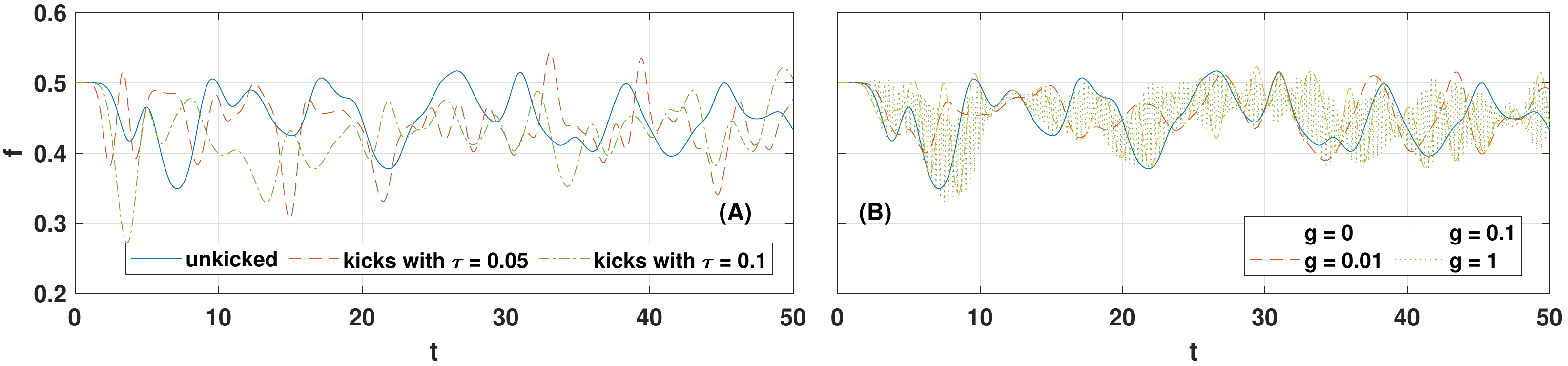}
 \caption{(A) Singlet fraction (f) vs time for the unkicked case is given by the blue plot and that with kicks is given by orange ($\tau = 0.05$) and green ($\tau = 0.1$) colored plots. There is a visible change in teleportation fidelity when we resort to the kicking scheme. For most of the time, $f$ is seen to decrease but we also see increased $f$ at a few instances of time. (B) Singlet fraction vs time for unkicked chains with varying degree of coupling with the environment ($g$). As the coupling constant is increased, the environment causes rapid oscillations in the singlet fraction as is evident from the graphs as well as Eq.~\ref{fiden}. All cases have been considered with chain length, $N=16$, environment chain length (in (B) only), $P=20$, $E_0=0.01$, $E_1=E_0/0.1=0.1$, $J_1=1$, $J_2=-1$.}
 \label{plot1}
\end{figure*}
\begin{figure*}
\includegraphics*[width=1\linewidth]{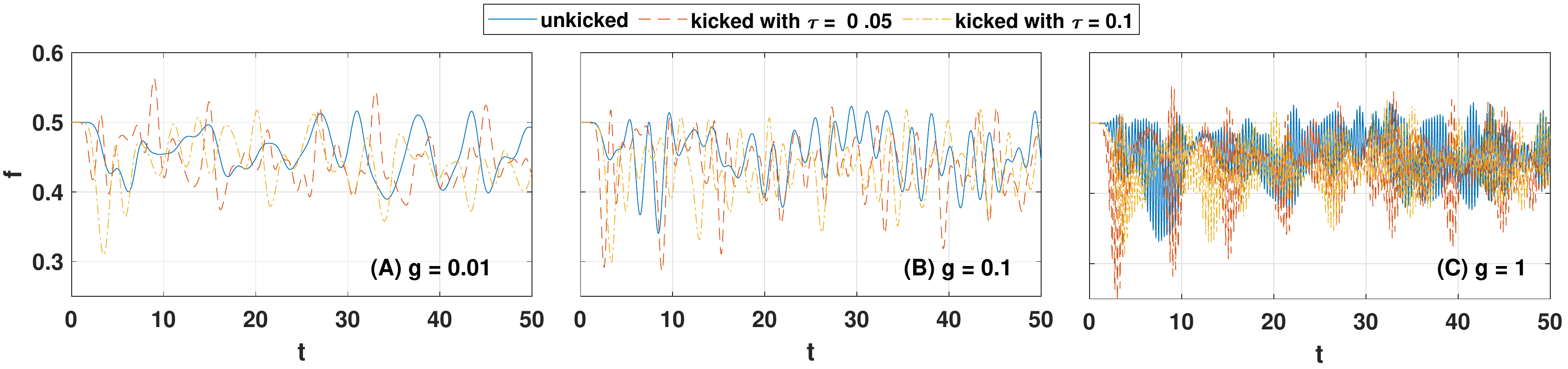}
 \caption{Singlet fraction vs time for kicked chains with varying degree of coupling with the environment in (A), (B) and (C) compared with the unkicked case. There are few instances of increase in $f$ due to kicking, but such instances are mitigated by the rapid oscillations caused by the environment in the strong coupling regime i.e. high $g$. However, a higher singlet fraction can still be obtained at some instances as shown in all cases. All cases have been considered with chain length, $N=16$, the number of environment sites $P=20$, $E_0=0.01$, $E_1=E_0/0.1=0.1$, $J_1=1$, $J_2=-1$.}
 \label{plot2}
\end{figure*}
\begin{figure*}[t]
\includegraphics*[width=1\linewidth]{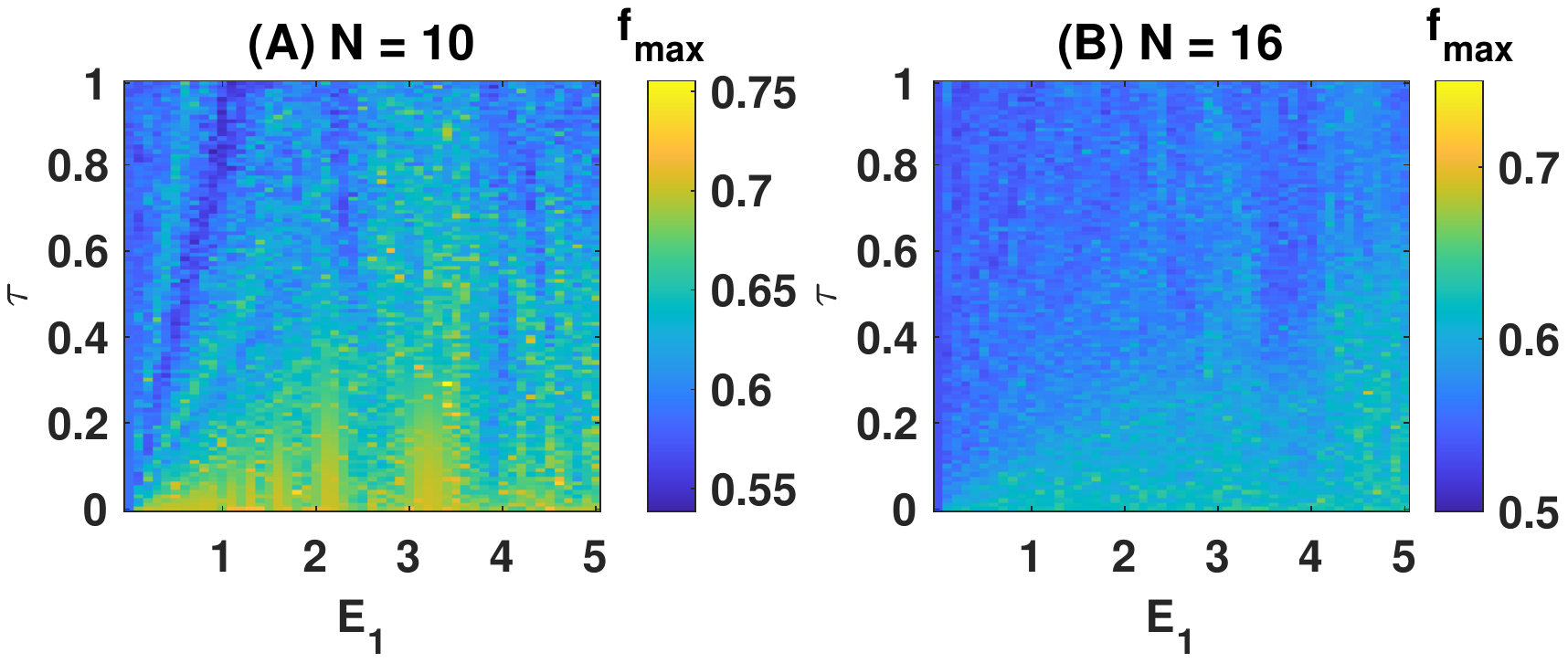}
\caption{$f_\textrm{max}$ which indicates that the singlet fraction is maximized over the number of kicks as a function of $\tau$ and $E_1$. $E_0=0.01$, $J_1=1$, $J_2=-1$ in all the cases. All values above 0.5 affirm the suitability of spin chain for entanglement sharing. For (A) $N=10$ and (B)$N=16$, the maximum value of singlet fraction 0.7552 and 0.6922, respectively, are achieved when maximized over $E_1$ and $\tau$.}
\label{tauvsE1}
\end{figure*}
\begin{figure*}[b]
\includegraphics*[width=1\linewidth]{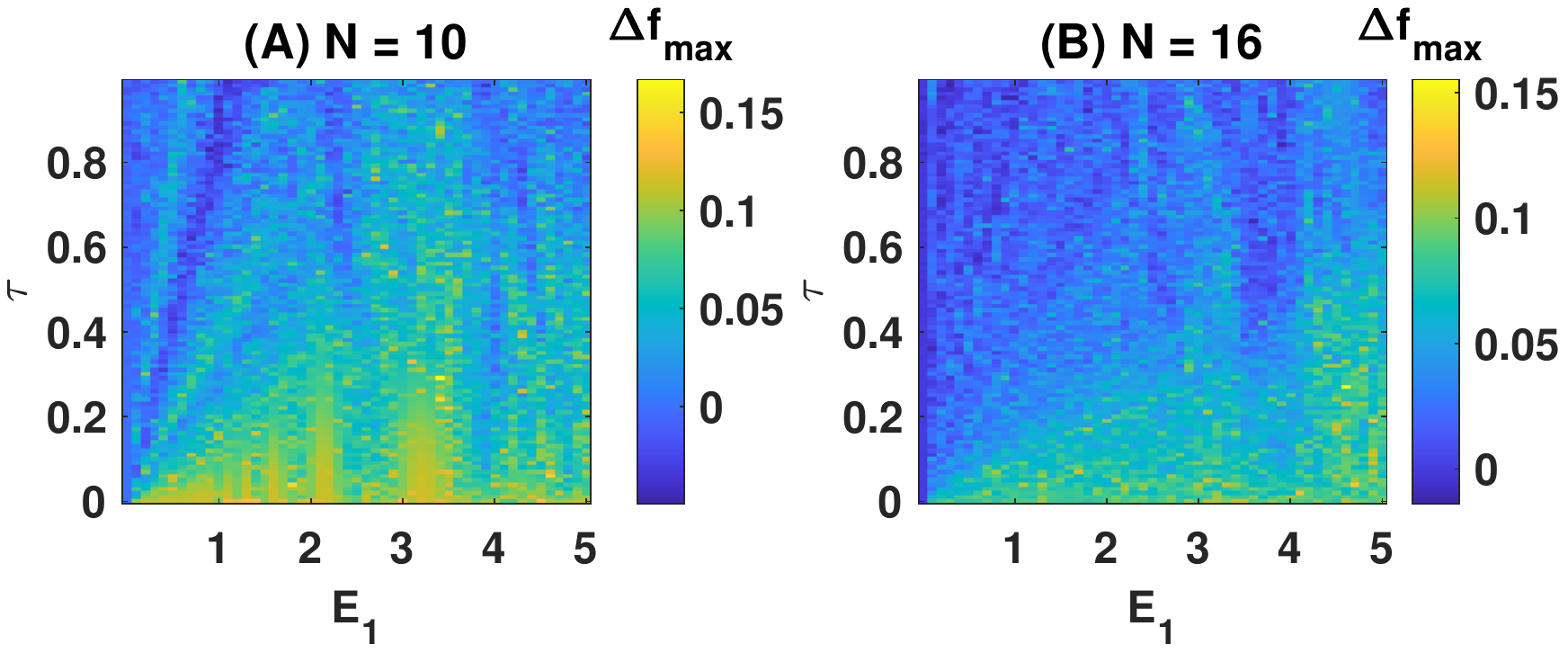}
\caption{Change in singlet fraction ($\Delta f_\textrm{max}$) obtained due to kicking for various chain lengths. The increase in singlet fraction is more significant for smaller chain (A) than longer chain (B). $E_0=0.01$, $J_1=1$, $J_2=-1$ in all the cases.}
\label{plot3}
\end{figure*}
\begin{figure*}
\includegraphics[width=1.0\linewidth]{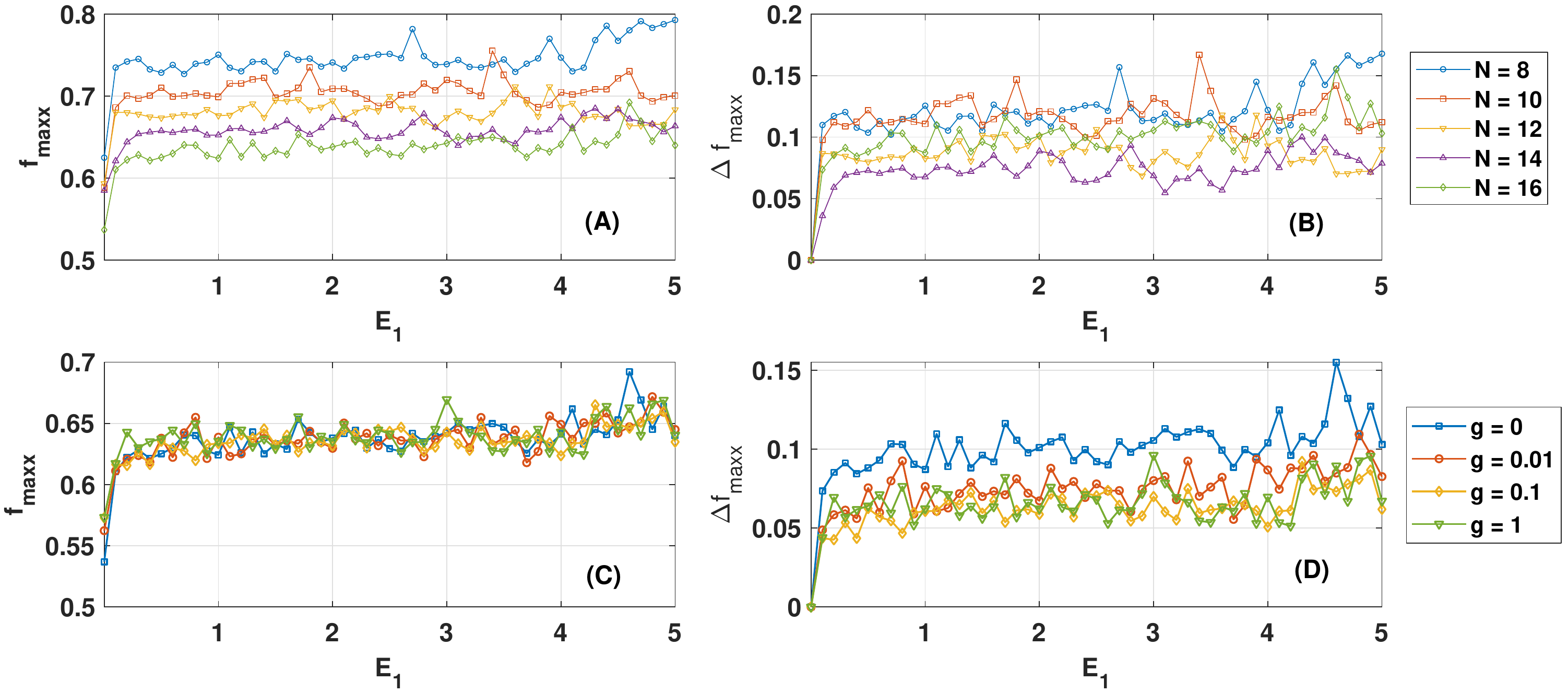}
 \caption{Maximum singlet fraction $f_\textrm{maxx}$ obtained for kicked chains maximized over the number of kicks and also the kick interval vs. the strength of the kicking electric field $E_1$ (A) for various chain lengths and (C) for various strengths of coupling between chain and environment ($g$) for fixed $N=16$, $P=20$. The corresponding change in maximum singlet fraction $\Delta f_{maxx}$ (B) for various chain lengths and (D) for various strengths of coupling between chain and environment ($g$) for fixed $N=16$, $P=20$. In all the cases, $E_0=0.01$, $J_1=1$, $J_2=-1$.}
 \label{plot4}
\end{figure*}
Using the protocol for sharing entanglement as outlined in the previous sections, we proceed to calculate the singlet fraction offered by the helical multiferroics and therefore, classify them as whether they are suitable for the purpose of being used in teleportation. We also probe the gain in singlet fraction offered by the kicking scheme. As expected from Eq.~\ref{f,sf}, teleportation fidelity would follow the same trend as the singlet fraction. 

We study the temporal characteristics of singlet fraction for a $\rm LiCu_2O_2$ spin chain for which, $J_2/J_1 \approx -1$. This has been shown for kicked and unkicked chains in Fig.~\ref{plot1} (A) with different kick intervals $\Delta \tau$. In Fig.~\ref{plot1} (B), we have shown the effect of environment on the temporal variation of singlet fraction for an unkicked chain. As expected from Eq.~\ref{fiden}, the environment is seen to cause oscillations in the singlet fraction which increase in frequency as the coupling constant between the environment and the chain is increased. Counter-intuitively, these oscillations due to the environment may also lead to higher singlet fraction than the case without the environment, especially in the case when the coupling -- $g$ is high. As we note from Fig.~\ref{plot1} (A), the kicked electric field offers enhanced singlet fraction in certain instances. Next, we have analyzed the temporal behaviour of singlet fraction in the presence of a kicked field and a uniform environment in Fig.~\eqref{plot2}. As before, certain instances show an increase in the singlet fraction. However, this may not be enough to mitigate the environmental effects at all the instances as seen from Fig.~\ref{plot2} (A), (B), and (C) which show this interplay for various values of the coupling -- $g$. The high oscillatory behaviour in the singlet fraction which is introduced by a strong coupling with the environment is evident from Fig.~\ref{plot2} (C).

Next, for the case of kicked chains, we have found the maximum singlet fraction ($f_\textrm{max}$) subject to varied number of kicks (i.e. the time of evolution) albeit bounded by the maximum time $t_\textrm{max}=1000$ and for the unkicked chains subject to varied time of evolution (also bounded by the maximum time $t_\textrm{max}=1000$). We have taken the difference between the maximum singlet fraction for kicked and unkicked cases i.e. $\Delta f_\textrm{max}= f_\textrm{max}^\textrm{kicked}-f_\textrm{max}^\textrm{unkicked}$ and shown its variation with the time interval between the kicks ($\tau$) and the amplitude of kicked electric field ($E_1$) in Fig.~\ref{plot3}. These parameters, alongwith the number of kicks required for $f_\textrm{max}$ are expected to be in our control through the duration,  frequency and shape of applied electric field \cite{kickrev} and hence, can be optimized to ensure high fidelity. Fig.~\ref{plot3} demonstrates that kicked electric field can be used to increase the singlet fraction. We observe that for $N=10$, $\Delta f_\textrm{max}$ increases more than that for $N=16$, especially at higher values of $E_1$, which is apparent from the color bars of the plots (A) and (B). Also, higher values of $\Delta f_\textrm{max}$ are reached more often for $N=10$ than $N=16$, due to the smaller length of the chain. For $N=16$, the maximum gain in singlet fraction due to kicking is $\sim 0.1$ which translates to a gain of $\sim 0.067$ in the teleportation fidelity. This is a $\sim 10\%$ increase for the fidelity in the range of $\sim0.6$.

Next, we introduce a new notation: $f_\textrm{maxx}$ for the maximum singlet fraction obtained after maximizing over the kick interval ($\tau$) as well. In this way, we ascertain the role of kicked electric field strength ($E_1$) on $f_\textrm{maxx}$ and $\Delta f_\textrm{maxx}$ in Fig.~\ref{plot4} (A) and (B) respectively albeit without any effect of the environment i.e. for $g=0$, and for varied lengths of the chain. We note that even the presence of small $E_1$ leads to a sharp increase in $\Delta f_\textrm{max}$ maximised over $\tau$. This behaviour is seen for all the chain lengths considered and is more for the shorter chain lengths. As seen from Fig.~\ref{plot4} (C), (D), even for different couplings between the environment and the spin chain (specifically shown for $N=16$), the electric field causes an increase in $\Delta f_\textrm{max}$ maximised over $\tau$ ($\equiv \Delta f_\textrm{maxx}$). This establishes that even in the presence of such an environment, the electric field can increase the singlet fraction and hence, the fidelity of teleportation. It is also noteworthy that the environment does not lead to a significant change in $\Delta f_{maxx}$ as illustrated in  Fig.~\ref{plot4} (C), (D) which is intuitive because it leads to rapid oscillations (as seen from Fig.~\eqref{plot1}) in the singlet fraction which more or less envelope trend without the environment. This translates into minimal increase/decrease when other parameters are optimized.

It is important to note here that we have only considered small lengths of spin chains due to exact diagonalization constraints. For such small chain lengths, in Fig.\eqref{plot4}(A), we see that $f_{maxx}$ decreases with the chain length which indicates that the entanglement sharing scheme may not be very effective for larger N as the singlet fraction would hover about 0.5.
 \section{Comparison with XX and XXZ models}
For the sake of completeness, we now compare the previous studies with our own, specifically the teleportation fidelities obtained in various other cases like \cite{Campos, venutixx, venuti,tele_arxiv}. In \cite{Campos}, a dimerized frustrated model has been considered which exhibits a near perfect transfer using a perfect singlet pair at the ends with end to end concurrence $=1$. Evidently, this has been achieved while optimizing the system parameters. In \cite{venutixx}, again perfect transfer has been shown to be possible by optimizing various parameters in a class of XX models. In \cite{venuti}, the system considered is antiferromagnetic Heisenberg chain with different couplings at the ends which has been shown to exhibit good teleportation fidelity at different finite temperatures. On a comparative scale, though our system fares a little poorer than the aforementioned instances, our model with some connections to a real material has an advantage of being susceptible to an external electric field, which we have shown to be useful for enhancing the teleportation fidelity. In any case, it is instructive to  study  and compare  various  models for entanglement sharing using  our protocol -- meaning by introducing a Bell pair at the middle of spin chain and expecting the system dynamics to avail appreciable entanglement at the receiver sites (i.e. the ends of the spin chain).
\begin{figure*}[t!]
\includegraphics[width=1\textwidth]{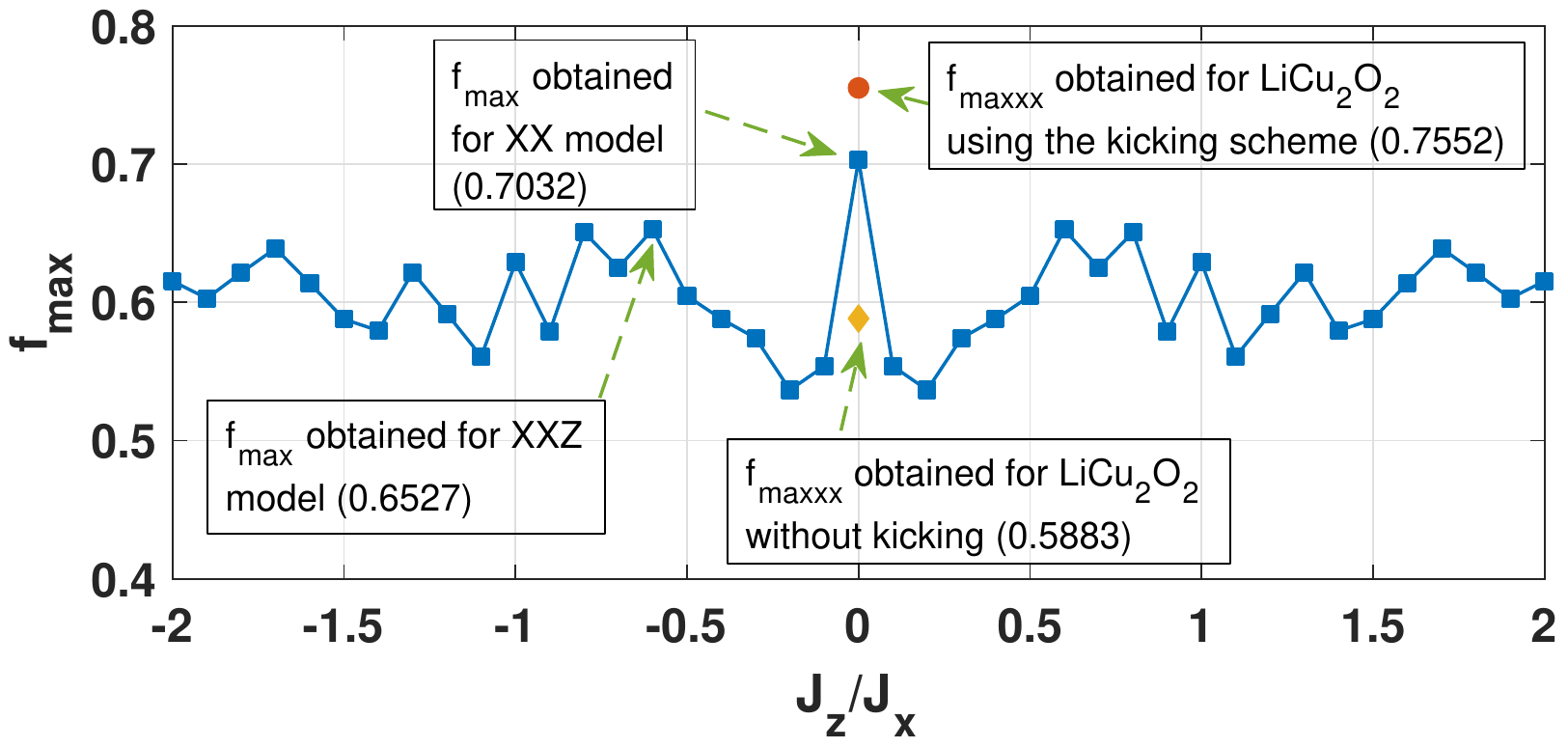}
 \caption{Singlet fraction (optimized over time of evolution bounded by $t_\textrm{max} = 1000$) $f_\textrm{max}$ vs $J_z/J_x$ in a class of XXZ models. The maximum singlet fraction obtained in the case of helical multiferroics after optimizing over the number of kicks, $\tau$ and $E_1 \in (0,5]$ and bounded by $t_\textrm{max} = 1000$ -- $f_\textrm{maxxx}$ has also been shown. All cases have been considered with chain length $N=10$. XXZ model has been considered for $J_x=1$ in all cases. Helical multiferroics have been considered with $E_0=0.01$, $J_1=1$, $J_2=-1$ in all the cases.}
 \label{plot10}
\end{figure*}

Fig.~\ref{plot10} shows that the XXZ model with $N=10$ exhibits a singlet fraction (maximized over time of evolution which is bounded by $t_\textrm{max} =1000$) $\sim0.6$ for various values of $J_z/J_x$ (Eq.~\ref{eq:hamXXZ}) which translates to a teleportation fidelity of $0.73$. As such, it fares comparable to most of the helical spin chains that we have considered. In the case when $J_z =0$, the XXZ model reduces to the XX model and the singlet fraction obtained is $0.7$. Therefore, the utility of XXZ and XX chains in our entanglement sharing protocol is established because $f_{max}>0.5$ . Now comparing the singlet fraction offered by helical multiferroics (optimized over the number of kicks, $\tau$ shown in Fig.~\ref{plot4} and also the kicked electric field $E_1 \in (0,5]$) of the same length ($N=10$) to that of XX, XXZ model, we find that if we select an optimum value of the kicked electric field ($E_1$), $LiCu_2O_2$ fares slightly better than both XX and XXZ chains, driven by an increase in the singlet fraction due to kicking.
\begin{figure*}
\includegraphics*[trim = 1cm 10cm 11cm 3.5cm,clip,width=1\linewidth]{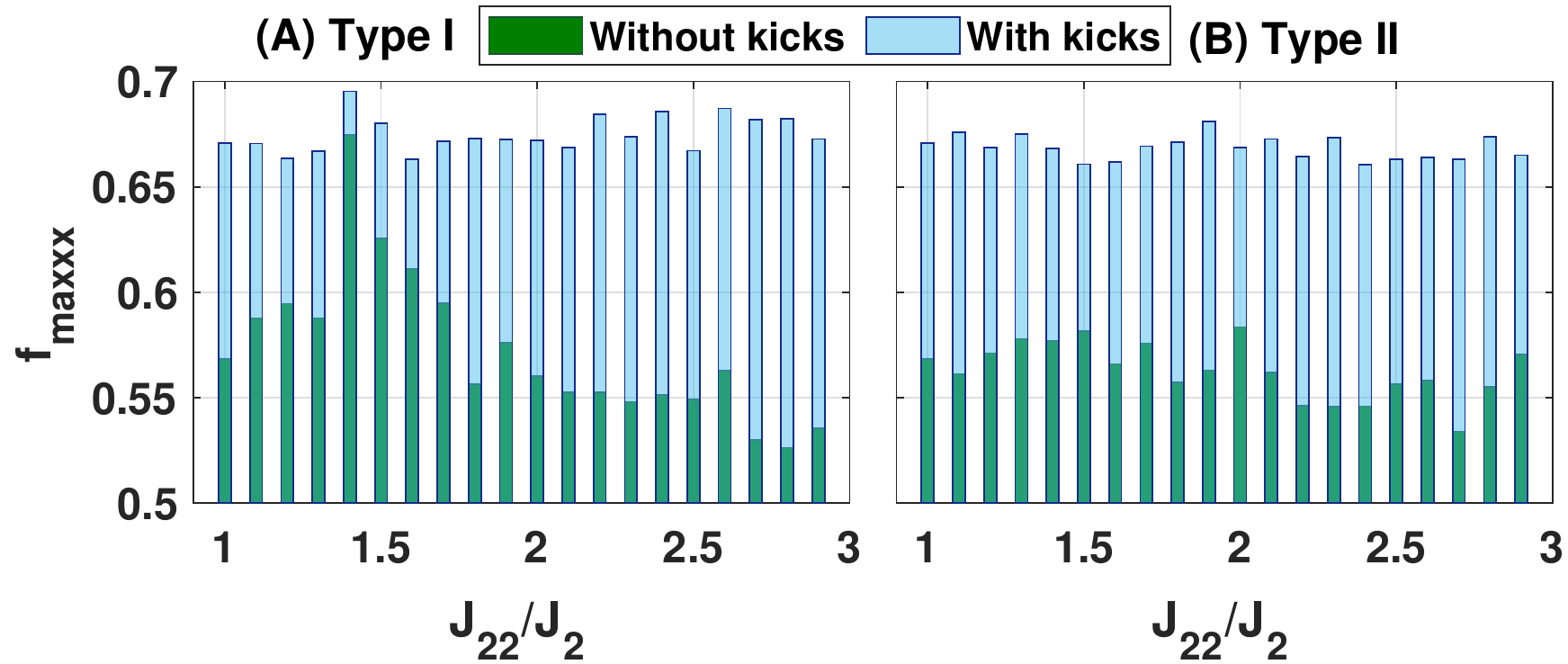}
 \caption{(A) Maximum singlet fraction ($f_\textrm{maxxx}$) vs impurity strength represented by $J_{22}/J_2$ for the unkicked case is given by the green bars and that with kicks is given by blue bars for embedded (A) Type I impurities (B) Type II impurities at sites 4 and 13 of the chain. There is a significant increment in singlet fraction when we resort to the kicking scheme. All the cases have been considered without the environment and with chain length, $N=16$,  $E_0=0.01$, $J_1=1$, $J_2=-1$. Introduction of Type I impurities is seen to drastically affect the singlet fraction as the impurity strength is increased. For $J_{22}/J_2 =1.4$ in (A) and $J_{22}/J_2 =1.9$ in (B), highest singlet fraction is obtained subject to optimum number of kicks, time interval between the kicks ($\tau$) and the kick strength ($E_1$).}
 \label{plot5}
\end{figure*}

\section{Effect of impurities} 
We now move on to ascertain the effect of the introduction of a specific impurities (defined as Type I in Fig.~\ref{fig:imp}) on the singlet fraction. We resort to ${\rm LiCu_2O_2}$ as the base system with $J_2/J_1=-1$ and change the values of exchange interaction strength $J_{22}/J_2$ which is representative of the increasing impurity strength in our case. As per the model considered, starting from $J_{11}/J_1 = J_{22}/J_2 = J_{222}/J_2 =1$, these exchange interaction strengths are tweaked in line with increment in the compression and hence, with the increasing strength of impurity. There are two impurities which are placed in both arms of the spin chain if taken from the middle of the chain at the sites 4 and 13. We first resort to Type I impurities and see the effect of their introduction in Fig.~\ref{plot5} (A) where the maximum singlet fraction ($f_\textrm{maxxx}$) is shown (maximized over number of kicks, time interval between the kicks ($\tau$, bounded by $t_\textrm{max}=1000$) and the kick strength ($E_1 \in (0,5]$). In the unkicked case, the optimization is just over the time of evolution which is also bounded by $t_\textrm{max}=1000$. Clearly, the effect of the introduction of impurities can be seen as the impurity strength (represented by $J_{22}/J_2$) is increased. The increase in strength of impurity leads to a considerable fall in the singlet fraction. However, the effect is much suppressed by the introduction of the kicking scheme. Even the chains characterised by high impurity strength which exhibited decreased singlet fraction now assume much higher values and the effect of impurity is largely mitigated at all impurity strengths. For the case of Type I impurities, highest singlet fraction is obtained for $J_{22}/J_2 = 1.4$ which is $f_\textrm{maxxx}=0.6953$. The corresponding teleportation fidelity is $f_\textrm{maxxx}=0.7969$. The other associated impurity parameters for this particular spin chain offering the highest singlet fraction are $J_{11}/J_2= 1.4$ and $J_{222}/J_2 = 0.7143$ ($\equiv 1/1.4$).

Similarly, we have also considered a case with two Type II impurities (Fig.~\ref{fig:imp}) embedded within the arms if taken from the middle of the spin chain at the sites 4 and 13. The base system remains the same i.e. $LiCu_2O_2$ and an increasing value of $J_{22}/J_2$ indicates a higher strength of impurity. As before, the other bond parameters relevant to the introduction of impurity i.e. $J_{111}$ and $J_{222}$ also change in proportion to the change in $J_{22}$. In Fig.~\ref{plot5} (B), we have considered the maximum singlet fraction obtainable without the kicks and compared it with the kicked case. As before, in the kicked case we have considered the maximum singlet fraction ($f_\textrm{maxxx}$) obtained by varied electric field strength $E_1 \in (0,5]$, number of kicks (bounded by $t_\textrm{max}=1000$), and time interval between the kicks ($\tau$). In the unkicked case, as before, the optimization is over the time of evolution which is also bounded by $t_\textrm{max}=1000$. From Fig.~\ref{plot5} (B), we see that the introduction of such impurities does not lead to considerable change in the singlet fraction as the impurity strength is increased which is in sharp contrast to the introduction of Type I impurities (apparent from Fig.~\ref{plot5} (A)). However, the kicking scheme still leads to the mitigation of the impurity affects and also an appreciable increase in the singlet fraction and hence, the teleportation fidelity. The highest singlet fraction of $f_\textrm{maxxx}=0.6812$ is obtained for $J_{22}/J_2 = 1.9$ for the kicked case. This translates to a teleportation fidelity of $f_\textrm{maxxx}=0.7875$. The other parameters associated with this impurity embedded system  are $J_{111}/J_2= 0.5263$ ($\equiv 1/1.9$) and $J_{222}/J_2 = 0.5263$ ($\equiv 1/1.9$).

\section{Conclusions}
The teleportation of quantum information is preceded by the distribution of entanglement between the parties involved. Teleportation fidelity quantifies the amount of shared entanglement available between the two distant parties. In this manuscript we explored the usage of an experimentally realizable helical multiferroic chain as an entanglement sharing channel.  
First, we calculated an expression for teleportation fidelity 
assuming a protocol where a Bell state ($\vert\Omega^{00}\rangle$) is prepared 
and introduced at the centre of a spin chain and entanglement at the end of 
chains is expected to be utilized in quantum teleportation. 
We have discovered 
the effectiveness of helical spin chains to be utilized in quantum teleportation by 
analyzing the teleportation fidelity. By resorting to a novel 
kicking scheme proposed in the manuscript, higher singlet fraction and 
teleportation fidelity may be obtained as shown in Fig.\eqref{plot4}, and Fig.\eqref{tauvsE1}. Most of the cases that we have studied 
already exhibited singlet fraction over $0.5$ which translates to
a teleportation fidelity of over $0.66$. However, with kicking, we have seen 
significant improvement over these values. This effect can be 
seen from Fig.~\ref{plot3} where the maximum singlet fraction subject to optimal
number of kicks and time interval has been plotted against the magnitude of kicked 
electric field ($E_1$). There is noticeable but irregular increase in the maximum singlet fraction obtainable.

We have also studied the effect of a common spin  environment to the multiferroic spin chain. This common spin environment (non self-interacting) is realized by placing the spin chain on the substrate of a longer one dimensional spin system with noninteracting spins. The uniform environment setting has been considered for derivation of the singlet fraction. Specifically, we have also probed strong and weak cases of coupling of the environment with the main system by changing the coupling constant $g$, though keeping it uniform at all the sites of the chain for simplicity. Surprisingly, our studies have shown that such a common spin environment can increase or decrease the maximal fidelity obtainable using a particular spin chain. This effect can be attributed to the oscillations caused by such an environment (Fig.~\ref{plot1} (B), Fig.~\ref{plot2}). This phenomenon is totally governed by the Eq.~\ref{fiden} though with added flexibility in the range of optimum kick interval $\tau$ and/or number of kicks. The interplay of all these factors may lead to a subtle increase/decrease in maximum teleportation fidelities which is hard to predict. Moreover, the previous works in this direction have also indicated counter-intuitive trends \cite{horo,bando,YY,ishi}. While using the kicking scheme, the optimal number of kicks and the kicking interval would ensure higher teleportation fidelity.

We compared the usefulness of various simpler spin chain models subject to our entanglement sharing protocol and do conclude that the results testify the usefulness of the helical, XX and XXZ spin chains for use in quantum teleportation as entanglement sharing channels using the protocol introduced in this manuscript. Moreover, the kicking scheme enables us to achieve higher fidelity than unkicked helical chains as well as XX and XXZ chains.

Finally we discussed the effects of the impurities and the novel kicking scheme. Though the introduction of a single impurity of either type has almost negligible effect on teleportation fidelity, the introduction of two such impurities (especially Type I) placed in both arms from the middle of spin chain has considerable effect notably when the impurity strengths are high. The kicking scheme mitigates the effect of the introduction of the impurities and prepares the otherwise unsuitable chains to be relevant for use in teleportation.

\begin{acknowledgements}
SKM acknowledges the Department of Science and Technology, India for support grant under the INSPIRE Faculty Fellowship award [IFA-12 PH 22]. JB and LC acknowledge the financial support of the DFG through SFB762.
\end{acknowledgements}

\printbibliography

@article{benn,
  title = {Teleporting an unknown quantum state via dual classical and Einstein-Podolsky-Rosen channels},
  author = {Bennett, Charles H. and Brassard, Gilles and Cr\'epeau, Claude and Jozsa, Richard and Peres, Asher and Wootters, William K.},
  journal = {Phys. Rev. Lett.},
  volume = {70},
  issue = {13},
  pages = {1895--1899},
  numpages = {0},
  year = {1993},
  publisher = {American Physical Society},
  doi = {10.1103/PhysRevLett.70.1895},
 url = {https://link.aps.org/doi/10.1103/PhysRevLett.70.1895}
}

@article{bose,
  title = {Quantum Communication through an Unmodulated Spin Chain},
  author = {Bose, Sougato},
  journal = {Phys. Rev. Lett.},
  volume = {91},
  issue = {20},
  pages = {207901},
  numpages = {4},
  year = {2003},
  publisher = {American Physical Society},
  doi = {10.1103/PhysRevLett.91.207901},
 url = {https://link.aps.org/doi/10.1103/PhysRevLett.91.207901}
}

@article{Wilde,
  title = {Strong and uniform convergence in the teleportation simulation of bosonic Gaussian channels},
  author = {Wilde, Mark M.},
  journal = {Phys. Rev. A},
  volume = {97},
  issue = {6},
  pages = {062305},
  numpages = {16},
  year = {2018},
  publisher = {American Physical Society},
  doi = {10.1103/PhysRevA.97.062305},
  url = {https://link.aps.org/doi/10.1103/PhysRevA.97.062305}
}

@article{Adesso,
  title = {Optimal Continuous Variable Quantum Teleportation with Limited Resources},
  author = {Liuzzo-Scorpo, Pietro and Mari, Andrea and Giovannetti, Vittorio and Adesso, Gerardo},
  journal = {Phys. Rev. Lett.},
  volume = {119},
  issue = {12},
  pages = {120503},
  numpages = {6},
  year = {2017},
  publisher = {American Physical Society},
  doi = {10.1103/PhysRevLett.119.120503},
  url = {https://link.aps.org/doi/10.1103/PhysRevLett.119.120503}
}

@article{Ivan,
  title = {All Entangled States can Demonstrate Nonclassical Teleportation},
  author = {Cavalcanti, Daniel and Skrzypczyk, Paul and \ifmmode \check{S}\else \v{S}\fi{}upi\ifmmode \acute{c}\else \'{c}\fi{}, Ivan},
  journal = {Phys. Rev. Lett.},
  volume = {119},
  issue = {11},
  pages = {110501},
  numpages = {5},
  year = {2017},
  publisher = {American Physical Society},
  doi = {10.1103/PhysRevLett.119.110501},
  url = {https://link.aps.org/doi/10.1103/PhysRevLett.119.110501}
}

@article{Rigolin,
  title = {Probabilistic quantum teleportation via thermal entanglement},
  author = {Fortes, Raphael and Rigolin, Gustavo},
  journal = {Phys. Rev. A},
  volume = {96},
  issue = {2},
  pages = {022315},
  numpages = {14},
  year = {2017},
  publisher = {American Physical Society},
  doi = {10.1103/PhysRevA.96.022315},
  url = {https://link.aps.org/doi/10.1103/PhysRevA.96.022315}
}

@article{Greplova,
  title = {Quantum teleportation with continuous measurements},
  author = {Greplova, Eliska and M\o{}lmer, Klaus and Andersen, Christian Kraglund},
  journal = {Phys. Rev. A},
  volume = {94},
  issue = {4},
  pages = {042334},
  numpages = {9},
  year = {2016},
  publisher = {American Physical Society},
  doi = {10.1103/PhysRevA.94.042334},
  url = {https://link.aps.org/doi/10.1103/PhysRevA.94.042334}
}

@article{Fortes,
  title = {Probabilistic quantum teleportation in the presence of noise},
  author = {Fortes, Raphael and Rigolin, Gustavo},
  journal = {Phys. Rev. A},
  volume = {93},
  issue = {6},
  pages = {062330},
  numpages = {9},
  year = {2016},
  publisher = {American Physical Society},
  doi = {10.1103/PhysRevA.93.062330},
  url = {https://link.aps.org/doi/10.1103/PhysRevA.93.062330}
}

@article{Campos,
  title = {Qubit Teleportation and Transfer across Antiferromagnetic Spin Chains},
  author = {Campos Venuti, L. and Degli Esposti Boschi, C. and Roncaglia, M.},
  journal = {Phys. Rev. Lett.},
  volume = {99},
  issue = {6},
  pages = {060401},
  numpages = {4},
  year = {2007},
  publisher = {American Physical Society},
  doi = {10.1103/PhysRevLett.99.060401},
  url = {https://link.aps.org/doi/10.1103/PhysRevLett.99.060401}
}

@article{venuti,
  title = {Long-Distance Entanglement in Spin Systems},
  author = {Campos Venuti, L. and Degli Esposti Boschi, C. and Roncaglia, M.},
  journal = {Phys. Rev. Lett.},
  volume = {96},
  issue = {24},
  pages = {247206},
  numpages = {4},
  year = {2006},
  publisher = {American Physical Society},
  doi = {10.1103/PhysRevLett.96.247206},
  url = {https://link.aps.org/doi/10.1103/PhysRevLett.96.247206}
}

@article{venutixx,
  title = {Long-distance entanglement and quantum teleportation in $XX$ spin chains},
  author = {Campos Venuti, L. and Giampaolo, S. M. and Illuminati, F. and Zanardi, P.},
  journal = {Phys. Rev. A},
  volume = {76},
  issue = {5},
  pages = {052328},
  numpages = {9},
  year = {2007},
  publisher = {American Physical Society},
  doi = {10.1103/PhysRevA.76.052328},
  url = {https://link.aps.org/doi/10.1103/PhysRevA.76.052328}
}

@article{ol,
  title = {Helical multiferroics for electric field controlled quantum information processing},
  author = {Azimi, M. and Chotorlishvili, L. and Mishra, S. K. and Greschner, S. and Vekua, T. and Berakdar, J.},
  journal = {Phys. Rev. B},
  volume = {89},
  issue = {2},
  pages = {024424},
  numpages = {8},
  year = {2014},
  publisher = {American Physical Society},
  doi = {10.1103/PhysRevB.89.024424},
  url = {https://link.aps.org/doi/10.1103/PhysRevB.89.024424}
}

@article{hv,
  title = {Qubit(s) transfer in helical spin chains},
  author = {Verma, Harshit and Chotorlishvili, L. and Berakdar, J. and Mishra, Sunil K.},
  journal = {EPL},
  volume = {119},
  issue = {3},
  pages= {30001},
  year = {2017},
  doi= {10.1209/0295-5075/119/30001},
  url= {https://doi.org/10.1209/0295-5075/119/30001},

}

@article{Mostovoy,
  title = {Ferroelectricity in Spiral Magnets},
  author = {Mostovoy, Maxim},
  journal = {Phys. Rev. Lett.},
  volume = {96},
  issue = {6},
  pages = {067601},
  numpages = {4},
  year = {2006},
  publisher = {American Physical Society},
  doi = {10.1103/PhysRevLett.96.067601},
  url = {https://link.aps.org/doi/10.1103/PhysRevLett.96.067601}
}

@article{Park,
  title = {Ferroelectricity in an $S=1/2$ Chain Cuprate},
  author = {Park, S. and Choi, Y. J. and Zhang, C. L. and Cheong, S-W.},
  journal = {Phys. Rev. Lett.},
  volume = {98},
  issue = {5},
  pages = {057601},
  numpages = {4},
  year = {2007},
  publisher = {American Physical Society},
  doi = {10.1103/PhysRevLett.98.057601},
  url = {https://link.aps.org/doi/10.1103/PhysRevLett.98.057601}
}

@article{Nagaosa,
  title = {Spin Current and Magnetoelectric Effect in Noncollinear Magnets},
  author = {Katsura, Hosho and Nagaosa, Naoto and Balatsky, Alexander V.},
  journal = {Phys. Rev. Lett.},
  volume = {95},
  issue = {5},
  pages = {057205},
  numpages = {4},
  year = {2005},
  publisher = {American Physical Society},
  doi = {10.1103/PhysRevLett.95.057205},
 url = {https://link.aps.org/doi/10.1103/PhysRevLett.95.057205}
}

@article{Chotorlishvili,
  title = {Superadiabatic quantum heat engine with a multiferroic working medium},
  author = {Chotorlishvili, L. and Azimi, M. and Stagraczy\ifmmode \acute{n}\else \'{n}\fi{}ski, S. and Toklikishvili, Z. and Sch\"uler, M. and Berakdar, J.},
  journal = {Phys. Rev. E},
  volume = {94},
  issue = {3},
  pages = {032116},
  numpages = {12},
  year = {2016},
  publisher = {American Physical Society},
  doi = {10.1103/PhysRevE.94.032116},
  url = {https://link.aps.org/doi/10.1103/PhysRevE.94.032116}
}

@article{Sekania,
  title = {Pulse and quench induced dynamical phase transition in a chiral multiferroic spin chain},
  author = {Azimi, M. and Sekania, M. and Mishra, S. K. and Chotorlishvili, L. and Toklikishvili, Z. and Berakdar, J.},
  journal = {Phys. Rev. B},
  volume = {94},
  issue = {6},
  pages = {064423},
  numpages = {12},
  year = {2016},
  publisher = {American Physical Society},
  doi = {10.1103/PhysRevB.94.064423},
  url = {https://link.aps.org/doi/10.1103/PhysRevB.94.064423}
}

@article{s5,
  title = {Multiferroic and magnetoelectric materials},
  author = {Eerenstein, W. and Mathur, N. D. and Scott, J. F.},
  journal = {Nature},
  volume = {442},
  pages = {749},
  numpages = {6},
  year = {2006},
  publisher = {Nature Publishing Group},
  doi = {10.1038/nature05023},
  url = {http://dx.doi.org/10.1038/nature05023}
}

@article {s6,
title = {The Renaissance of Magnetoelectric Multiferroics},
author = {Spaldin, Nicola A. and Fiebig, Manfred},
journal = {Science},
volume = {309},
issue = {5733},
pages = {391--392},
numpages = {2},
year = {2005},
publisher = {American Association for the Advancement of Science},
doi = {10.1126/science.1113357},
url = {http://science.sciencemag.org/content/309/5733/391},
}

@article{s7,
  title = {Multiferroics: a magnetic twist for ferroelectricity},
  author = {Cheong, Sang-Wook and Mostovoy, Maxim},
  journal = {Nature Materials},
  volume = {6},
  pages = {13},
  numpages = {7},
  year = {2007},
  publisher = {Nature Publishing Group},
  doi = {10.1038/nmat1804},
 url = {http://dx.doi.org/10.1038/nmat1804}
}

@article{Oh2002,
  title = {Multipartite entanglement for entanglement teleportation},
  author = {Lee, Jinhyoung and Min, Hyegeun and Oh, Sung Dahm},
  journal = {Phys. Rev. A},
  volume = {66},
  issue = {5},
  pages = {052318},
  numpages = {5},
  year = {2002},
  publisher = {American Physical Society},
  doi = {10.1103/PhysRevA.66.052318},
  url = {https://link.aps.org/doi/10.1103/PhysRevA.66.052318}
}

@article{horo,
  title = {Local environment can enhance fidelity of quantum teleportation},
  author = {Badzia\ifmmode \mbox{\c{}}\else \c{}\fi{}g, Piotr and Horodecki, Micha\l{} and Horodecki, Pawe\l{} and Horodecki, Ryszard},
  journal = {Phys. Rev. A},
  volume = {62},
  issue = {1},
  pages = {012311},
  numpages = {7},
  year = {2000},
  publisher = {American Physical Society},
  doi = {10.1103/PhysRevA.62.012311},
  url = {https://link.aps.org/doi/10.1103/PhysRevA.62.012311}
}

@article{bando,
  title = {Origin of noisy states whose teleportation fidelity can be enhanced through dissipation},
  author = {Bandyopadhyay, Somshubhro},
  journal = {Phys. Rev. A},
  volume = {65},
  issue = {2},
  pages = {022302},
  numpages = {6},
  year = {2002},
  publisher = {American Physical Society},
  doi = {10.1103/PhysRevA.65.022302},
  url = {https://link.aps.org/doi/10.1103/PhysRevA.65.022302}
}

@article{ishi,
  title = {Quantum channel locally interacting with environment},
  author = {Ishizaka, Satoshi},
  journal = {Phys. Rev. A},
  volume = {63},
  issue = {3},
  pages = {034301},
  numpages = {4},
  year = {2001},
  publisher = {American Physical Society},
  doi = {10.1103/PhysRevA.63.034301},
  url = {https://link.aps.org/doi/10.1103/PhysRevA.63.034301}
}

@article{cuch,
  title = {Decoherence from spin environments},
  author = {Cucchietti, F. M. and Paz, J. P. and Zurek, W. H.},
  journal = {Phys. Rev. A},
  volume = {72},
  issue = {5},
  pages = {052113},
  numpages = {8},
  year = {2005},
  publisher = {American Physical Society},
  doi = {10.1103/PhysRevA.72.052113},
  url = {https://link.aps.org/doi/10.1103/PhysRevA.72.052113}
}

@article{Cai,
  title = {Decoherence effects on the quantum spin channels},
  author = {Cai, Jian-Ming and Zhou, Zheng-Wei and Guo, Guang-Can},
  journal = {Phys. Rev. A},
  volume = {74},
  issue = {2},
  pages = {022328},
  numpages = {6},
  year = {2006},
  publisher = {American Physical Society},
  doi = {10.1103/PhysRevA.74.022328},
  url = {https://link.aps.org/doi/10.1103/PhysRevA.74.022328}
}

@article{Giampaolo,
  title = {Long-distance entanglement and quantum teleportation in coupled-cavity arrays},
  author = {Giampaolo, Salvatore M. and Illuminati, Fabrizio},
  journal = {Phys. Rev. A},
  volume = {80},
  issue = {5},
  pages = {050301},
  numpages = {4},
  year = {2009},
  publisher = {American Physical Society},
  doi = {10.1103/PhysRevA.80.050301},
  url = {https://link.aps.org/doi/10.1103/PhysRevA.80.050301}
}

@article{Horodecki,
  title = {General teleportation channel, singlet fraction, and quasidistillation},
  author = {Horodecki, Micha\l{} and Horodecki, Pawe\l{} and Horodecki, Ryszard},
  journal = {Phys. Rev. A},
  volume = {60},
  issue = {3},
  pages = {1888--1898},
  numpages = {10},
  year = {1999},
  publisher = {American Physical Society},
  doi = {10.1103/PhysRevA.60.1888},
  url = {https://link.aps.org/doi/10.1103/PhysRevA.60.1888}
}

@article{Li2013,
title = {Entanglement fidelity of the standard quantum teleportation channel},
author = {Gang Li and Ming-Yong Ye and Xiu-Min Lin},
journal = {Physics Letters A},
volume = {377},
issue= {23},
numpages= {3},
pages = {1531 - 1533},
year = {2013},
publisher = {Elsevier},
doi = {10.1016/j.physleta.2013.04.034},
url = {http://www.sciencedirect.com/science/article/pii/S0375960113004131}
}

@article{Albanese2004,
  title = {Mirror Inversion of Quantum States in Linear Registers},
  author = {Albanese, Claudio and Christandl, Matthias and Datta, Nilanjana and Ekert, Artur},
  journal = {Phys. Rev. Lett.},
  volume = {93},
  issue = {23},
  pages = {230502},
  numpages = {4},
  year = {2004},
  publisher = {American Physical Society},
  doi = {10.1103/PhysRevLett.93.230502},
  url = {https://link.aps.org/doi/10.1103/PhysRevLett.93.230502}
}

@article{Christandl2005,
  title = {Perfect transfer of arbitrary states in quantum spin networks},
  author = {Christandl, Matthias and Datta, Nilanjana and Dorlas, Tony C. and Ekert, Artur and Kay, Alastair and Landahl, Andrew J.},
  journal = {Phys. Rev. A},
  volume = {71},
  issue = {3},
  pages = {032312},
  numpages = {11},
  year = {2005},
  publisher = {American Physical Society},
  doi = {10.1103/PhysRevA.71.032312},
  url = {https://link.aps.org/doi/10.1103/PhysRevA.71.032312}
}

@article{Boness2006,
  title = {Entanglement and Dynamics of Spin Chains in Periodically Pulsed Magnetic Fields: Accelerator Modes},
  author = {Boness, T. and Bose, S. and Monteiro, T. S.},
  journal = {Phys. Rev. Lett.},
  volume = {96},
  issue = {18},
  pages = {187201},
  numpages = {4},
  year = {2006},
  publisher = {American Physical Society},
  doi = {10.1103/PhysRevLett.96.187201},
  url = {https://link.aps.org/doi/10.1103/PhysRevLett.96.187201}
}

@article{Banchi2010,
  title = {Optimal dynamics for quantum-state and entanglement transfer through homogeneous quantum systems},
  author = {Banchi, L. and Apollaro, T. J. G. and Cuccoli, A. and Vaia, R. and Verrucchi, P.},
  journal = {Phys. Rev. A},
  volume = {82},
  issue = {5},
  pages = {052321},
  numpages = {5},
  year = {2010},
  publisher = {American Physical Society},
  doi = {10.1103/PhysRevA.82.052321},
  url = {https://link.aps.org/doi/10.1103/PhysRevA.82.052321}
}

@article{Apollaro2012,
  title = {Fidelity ballistic quantum-state transfer through long uniform channels},
  author = {Apollaro, T. J. G. and Banchi, L. and Cuccoli, A. and Vaia, R. and Verrucchi, P.},
  journal = {Phys. Rev. A},
  volume = {85},
  issue = {5},
  pages = {052319},
  numpages = {11},
  year = {2012},
  publisher = {American Physical Society},
  doi = {10.1103/PhysRevA.85.052319},
  url = {https://link.aps.org/doi/10.1103/PhysRevA.85.052319}
}

@article{Schrettle2008,
  title = {Switching the ferroelectric polarization in the S=1∕2 chain cuprate $\mathrm{Li}\mathrm{Cu}\mathrm{V}{\mathrm{O}}_{4}$ by external magnetic fields},
  author = {Schrettle, F. and Krohns, S. and Lunkenheimer, P. and Hemberger, J. and B\"uttgen, N. and Krug von Nidda, H.-A. and Prokofiev, A. V. and Loidl, A.},
  journal = {Phys. Rev. B},
  volume = {77},
  issue = {14},
  pages = {144101},
  numpages = {6},
  year = {2008},
  publisher = {American Physical Society},
  doi = {10.1103/PhysRevB.77.144101},
  url = {https://link.aps.org/doi/10.1103/PhysRevB.77.144101}
}

@article{Menzel2012,
  title = {Information Transfer by Vector Spin Chirality in Finite Magnetic Chains},
  author = {Menzel, Matthias and Mokrousov, Yuriy and Wieser, Robert and Bickel, Jessica E. and Vedmedenko, Elena and Bl\"ugel, Stefan and Heinze, Stefan and von Bergmann, Kirsten and Kubetzka, Andr\'e and Wiesendanger, Roland},
  journal = {Phys. Rev. Lett.},
  volume = {108},
  issue = {19},
  pages = {197204},
  numpages = {5},
  year = {2012},
  publisher = {American Physical Society},
  doi = {10.1103/PhysRevLett.108.197204},
  url = {https://link.aps.org/doi/10.1103/PhysRevLett.108.197204}
}

@article{kickrev,
title = "Charge and spin dynamics driven by ultrashort extreme broadband pulses: A theory perspective",
journal = "Physics Reports",
volume = "672",
pages = "1 - 82",
year = "2017",
issn = "0370-1573",
doi = "https://doi.org/10.1016/j.physrep.2016.12.005",
url = "http://www.sciencedirect.com/science/article/pii/S0370157317300017",
author = "Andrey S. Moskalenko and Zhen-Gang Zhu and Jamal Berakdar",
}

@article{Yeo_2005,
	doi = {10.1088/0305-4470/38/14/012},
	url = {https://doi.org/10.1088/0305-4470/38/14/012},
	year = 2005,
	publisher = {{IOP} Publishing},
	volume = {38},
	number = {14},
	pages = {3235--3243},
	author = {Ye Yeo and Tongqi Liu and Yu-En Lu and Qi-Zhong Yang},
	title = {Quantum teleportation via a two-qubit {HeisenbergXYchain}{\textemdash}effects of anisotropy and magnetic field},
	journal = {Journal of Physics A: Mathematical and General}
}

@article{YY,
  title = {Local noise can enhance two-qubit teleportation},
  author = {Yeo, Ye},
  journal = {Phys. Rev. A},
  volume = {78},
  issue = {2},
  pages = {022334},
  numpages = {7},
  year = {2008},
  publisher = {American Physical Society},
  doi = {10.1103/PhysRevA.78.022334},
  url = {https://link.aps.org/doi/10.1103/PhysRevA.78.022334}
}

@article{Yeo_2009,
	doi = {10.1209/0295-5075/86/40009},
	url = {https://doi.org/10.1209/0295-5075/86/40009},
	year = 2009,
	publisher = {{IOP} Publishing},
	volume = {86},
	number = {4},
	pages = {40009},
	author = {Ye Yeo and Zhe-Wei Kho and Lixian Wang},
	title = {Effects of Pauli channels and noisy quantum operations on standard teleportation},
	journal = {{EPL} (Europhysics Letters)}
}

@article{tele_arxiv,
  title = {Spin chains for two-qubit teleportation},
  author = {Apollaro, Tony J. G. and Almeida, Guilherme M. A. and Lorenzo, Salvatore and Ferraro, Alessandro and Paganelli, Simone},
  journal = {Phys. Rev. A},
  volume = {100},
  issue = {5},
  pages = {052308},
  numpages = {8},
  year = {2019},
  publisher = {American Physical Society},
  doi = {10.1103/PhysRevA.100.052308},
  url = {https://link.aps.org/doi/10.1103/PhysRevA.100.052308}
}

@article {Zajac439,
	author = {Zajac, D. M. and Sigillito, A. J. and Russ, M. and Borjans, F. and Taylor, J. M. and Burkard, G. and Petta, J. R.},
	title = {Resonantly driven CNOT gate for electron spins},
	volume = {359},
	number = {6374},
	pages = {439--442},
	year = {2018},
	doi = {10.1126/science.aao5965},
	publisher = {American Association for the Advancement of Science},
	URL = {https://science.sciencemag.org/content/359/6374/439},
	eprint = {https://science.sciencemag.org/content/359/6374/439.full.pdf},
	journal = {Science}
}

@article{rout1,
  title = {Quantum router based on ac control of qubit chains},
  author = {Zueco, David and Galve, Fernando and Kohler, Sigmund and H\"anggi, Peter},
  journal = {Phys. Rev. A},
  volume = {80},
  issue = {4},
  pages = {042303},
  numpages = {10},
  year = {2009},
  publisher = {American Physical Society},
  doi = {10.1103/PhysRevA.80.042303},
  url = {https://link.aps.org/doi/10.1103/PhysRevA.80.042303}
}

@article{rout2,
  title = {Entanglement Routers Using Macroscopic Singlets},
  author = {Bayat, Abolfazl and Bose, Sougato and Sodano, Pasquale},
  journal = {Phys. Rev. Lett.},
  volume = {105},
  issue = {18},
  pages = {187204},
  numpages = {4},
  year = {2010},
  publisher = {American Physical Society},
  doi = {10.1103/PhysRevLett.105.187204},
  url = {https://link.aps.org/doi/10.1103/PhysRevLett.105.187204}
}

@article{rout3,
  title = {Information-transferring ability of the different phases of a finite XXZ spin chain},
  author = {Bayat, Abolfazl and Bose, Sougato},
  journal = {Phys. Rev. A},
  volume = {81},
  issue = {1},
  pages = {012304},
  numpages = {11},
  year = {2010},
  publisher = {American Physical Society},
  doi = {10.1103/PhysRevA.81.012304},
  url = {https://link.aps.org/doi/10.1103/PhysRevA.81.012304}
}

@article{rout4,
  title = {Routing quantum information in spin chains},
  author = {Paganelli, Simone and Lorenzo, Salvatore and Apollaro, Tony J. G. and Plastina, Francesco and Giorgi, Gian Luca},
  journal = {Phys. Rev. A},
  volume = {87},
  issue = {6},
  pages = {062309},
  numpages = {8},
  year = {2013},
  publisher = {American Physical Society},
  doi = {10.1103/PhysRevA.87.062309},
  url = {https://link.aps.org/doi/10.1103/PhysRevA.87.062309}
}

@article{Hu2009,
author={Hu, M. L.
and Lian, H. L.},
title={State transfer in intrinsic decoherence spin channels},
journal={The European Physical Journal D},
year={2009},
month={Aug},
day={07},
volume={55},
number={3},
pages={711},
issn={1434-6079},
doi={10.1140/epjd/e2009-00220-8},
url={https://doi.org/10.1140/epjd/e2009-00220-8}
}

\end{document}